\documentclass[final,authoryear,5p,times,twocolumn]{elsarticle}

\usepackage{epsfig}

\usepackage{amssymb}

\usepackage{lineno}
\usepackage{rotating}
\usepackage[bookmarks = true, bookmarksnumbered = true, pdfpagemode =None, pdfstartview = FitH, pdfpagelayout = SinglePage, colorlinks = true, urlcolor = blue, citecolor = monbleu]{hyperref}

\newcommand{\pderiv}[2]{\frac{\partial #1}{\partial #2}}

\newcommand{\beq}{\bigskip\begin{equation}}
\newcommand{\eeq}{\bigskip\end{equation}}

\newcommand{\m}{$\mu$m}

\journal{Icarus}

\begin{document}

\begin{frontmatter}



\title{Seasonal Evolution of Saturn's Polar Temperatures and Composition}


\author[ox]{Leigh N. Fletcher}
\ead{fletcher@atm.ox.ac.uk}
\author[ox]{P.G.J. Irwin}
\author[ox]{J.A. Sinclair}
\author[jpl]{G.S. Orton}
\author[ox]{R.S. Giles}
\author[ral]{J. Hurley}
\author[cu]{N. Gorius}
\author[umd]{R.K. Achterberg}
\author[umd]{B.E. Hesman}
\author[gsfc]{G.L. Bjoraker}

\address[ox]{Atmospheric, Oceanic \& Planetary Physics, Department of Physics, University of Oxford, Clarendon Laboratory, Parks Road, Oxford, OX1 3PU, UK}
\address[jpl]{Jet Propulsion Laboratory, California Institute of Technology, 4800 Oak Grove Drive, Pasadena, CA, 91109, USA}
\address[ral]{STFC Rutherford Appleton Laboratory, Harwell Science and Innovation Campus, Didcot, OX11 0QX, UK}
\address[cu]{Department of Physics, The Catholic University of America, Washington, DC 20064, USA}
\address[umd]{Department of Astronomy, University of Maryland, College Park, MD 20742, USA.}
\address[gsfc]{NASA/Goddard Space Flight Center, Greenbelt, Maryland, 20771, USA }

\begin{abstract}

The seasonal evolution of Saturn's polar atmospheric temperatures and hydrocarbon composition is derived from a decade of Cassini Composite Infrared Spectrometer (CIRS) 7-16 $\mu$m thermal infrared spectroscopy.  We construct a near-continuous record of atmospheric variability poleward of 60$^\circ$  from northern winter/southern summer (2004, $L_s=293^\circ$) through the equinox (2009, $L_s=0^\circ$) to northern spring/southern autumn (2014, $L_s=56^\circ$).  The hot tropospheric polar cyclones that are entrained by prograde jets within 2-3$^\circ$ of each pole, and the hexagonal shape of the north polar belt, are both persistent features throughout the decade of observations.  The hexagon vertices rotated westward by $\approx30^\circ$ longitude between March 2007 and April 2013, confirming that they are not stationary in the Voyager-defined System III longitude system as previously thought.  Tropospheric temperature contrasts between the cool polar zones (near 80-85$^\circ$) and warm polar belts (near 75-80$^\circ$) have varied in both hemispheres, resulting in changes to the vertical windshear on the zonal jets in the upper troposphere and lower stratosphere.  The extended region of south polar stratospheric emission has cooled dramatically poleward of the sharp temperature gradient near 75$^\circ$S (by approximately -5 K/yr), coinciding with a depletion in the abundances of acetylene ($0.030\pm0.005$ ppm/yr) and ethane ($0.35\pm0.1$ ppm/yr), and suggestive of stratospheric upwelling with vertical wind speeds of $w\approx+0.1$ mm/s.  The upwelling appears most intense within 5$^\circ$ latitude of the south pole. This is mirrored by a general warming of the northern polar stratosphere (+5 K/yr) and an enhancement in acetylene ($0.030\pm0.003$ ppm/yr) and ethane ($0.45\pm0.1$ ppm/yr) abundances that appears to be most intense poleward of 75$^\circ$N, suggesting subsidence at $w\approx-0.15$ mm/s.  However, the sharp gradient in stratospheric emission expected to form near 75$^\circ$N by northern summer solstice (2017, $L_s=90^\circ$) has not yet been observed, so we continue to await the development of a northern summer stratospheric vortex.  The peak stratospheric warming in the north occurs at lower pressure levels ($p<1$ mbar) than the peak stratospheric cooling in the south ($p>1$ mbar).  Vertical motions are derived from both the temperature field (using the measured rates of temperature change and the deviations from the expectations of radiative equilibrium models) and hydrocarbon distributions (solving the continuity equation).  Vertical velocities tend towards zero in the upper troposphere where seasonal temperature contrasts are smaller, except within the tropospheric polar cyclones where $w\approx\pm0.02$ mm/s.  North polar minima in tropospheric and stratospheric temperatures were detected in 2008-2010 (lagging one season, or 6-8 years, behind winter solstice); south polar maxima appear to have occurred before the start of the Cassini observations (1-2 years after summer solstice), consistent with the expectations of radiative climate models.  The influence of dynamics implies that the coldest winter temperatures occur in the $75-80^\circ$ region in the stratosphere, and in the cool polar zones in the troposphere, rather than at the poles themselves.  In addition to vertical motions, we propose that the UV-absorbent polar stratospheric aerosols entrained within Saturn's vortices contribute significantly to the radiative budget at the poles, adding to the localised enhancement in the south polar cooling and north polar warming poleward of $\pm75^\circ$.  

\end{abstract}


\end{frontmatter}

\section{Introduction}
\label{intro}

The polar regions of the giant planets exhibit some of the most complex environmental conditions found in the outer solar system.  They are the apex of a planet-wide circulation system, where the organised pattern of zonal banding gives way to the more mottled appearance of the high latitudes.  Furthermore, they are the site of the closest connection between the neutral atmosphere and the charged magnetosphere via auroral activity, and the unusual chemistry and aerosol production generated by this energy injection.  Given Saturn's 26.7$^\circ$ obliquity, its poles are subjected to extremes of insolation over its 29.5-year orbit, spending a decade in polar night before emerging into spring sunlight.  As the polar winter is hidden from Earth-based observers, only an orbiting spacecraft can provide the vantage point and longevity to study the evolving polar atmosphere as the seasons change.  After a decade of exploration (2004-2014, planetocentric solar longitudes of $L_s=293-56^\circ$), Cassini has provided the most comprehensive view of a seasonally-evolving giant planet ever obtained.  This study focuses on temporal evolution at Saturn's poles as southern summer became southern autumn, and northern winter became northern spring.  

This investigation builds upon the snapshot of polar conditions observed in 2005-2007 \citep{08fletcher_poles} in the thermal infrared (7-1000 \m) by the Cassini Composite Infrared Spectrometer \citep[CIRS,][]{04flasar}.  Those observations revealed a striking asymmetry in temperature from the summer to the winter pole, with an extended warm stratospheric `hood' over the summer pole suggesting the presence of a summer stratospheric vortex (approximately $75-90^\circ$S planetographic latitude), possibly entraining unique polar aerosols, that was absent from the cold winter pole.  In addition, compact and warm cyclonic vortices were discovered within $2-3^\circ$ of both poles \citep{05orton, 08fletcher_poles}, suggesting that these cyclonic polar vortices \citep[and their `hurricane-like' eyewalls,][]{06sanchez,08dyudina,09dyudina,09baines_pole} are persistent features on Saturn, irrespective of the season.  We refer to these as the `North Polar Cyclone' (NPC) and `South Polar Cyclone' (SPC), respectively.  We use a decade of Cassini observations to study the stability of the polar cyclones, the seasonal evolution of the summer stratospheric vortices, and to search for evidence of compositional trends in the polar regions.  

Previous studies of Saturn's evolving atmospheric structure \citep{10fletcher_seasons, 10li, 11guerlet, 13sinclair, 13li,14sinclair} have used nadir and limb observations acquired while Cassini was in a near-equatorial orbit, making observations of the highest latitudes difficult.  \citet{10fletcher_seasons} discovered a cooling of the summer stratosphere over a five-year period in pre-equinox (August 2009) CIRS spectra, consistent with the expectations of radiative cooling \citep{08greathouse_agu}, which suggested a weakening of the peripheral stratospheric jet entraining the vortex.  This southern cooling was observed through to 2010 by \citet{13sinclair}. A minimum in the northern stratospheric temperatures was observed at $78-82^\circ$N, the latitude of the hexagonal wave in Saturn's troposphere \citep{88godfrey}.  Poleward of the hexagon, the north pole was warmer than expected due to subsidence and adiabatic heating, despite the winter conditions.


The timescale for the formation of the summer stratospheric vortex was not constrained by observations prior to Cassini's arrival.  Mid-infrared imaging from Keck in February 2004 ($L_s=287.4^\circ$) showed that the elevated south polar emission was already present \citep{05orton}, and must have developed prior to southern summer solstice (October 2002, $L_s=270^\circ$, G. Orton, \textit{pers. comms.}).  Intriguingly, stratosphere-sensitive imaging from NASA's Infrared Telescope Facility in March 1989 ($L_s=104.5^\circ$, after northern summer solstice) \citep{89gezari} showed enhanced emission from the north polar region, suggesting that the Cassini mission should expect the onset of a warm northern summer stratospheric vortex between now and the northern summer solstice in May 2017 ($L_s=90^\circ$).  This study aims to constrain the timescale for the seasonal onset of the northern stratospheric vortex and the dissipation of the southern stratospheric vortex.  

In addition to these thermal changes, \citet{13sinclair} used two epochs of low-spectral resolution CIRS data ($\Delta\nu=15$ cm$^{-1}$ apodised) acquired from low-inclination orbits to demonstrate that both ethane (C$_2$H$_6$) and acetylene (C$_2$H$_2$) were enhanced at the south pole but had decreased in concentration between 2005 and 2010.  In this work we significantly extend the temporal and spatial coverage of the CIRS dataset and the robustness of the spectral inversions by considering all available data at $\Delta\nu=2.5$ cm$^{-1}$ and $\Delta\nu=15.0$ cm$^{-1}$ spectral resolution, particularly those acquired during Cassini's inclined orbits, as described in Section \ref{obs}.  The methods used to invert these spectra are described in Section \ref{method}, and the resulting temperature and composition variability is presented in Section \ref{results}.  The results are compared with the expectations of seasonal climate models \citep[e.g.,][]{08greathouse_agu, 12friedson, 14guerlet}, with implications for horizontal and vertical motions in Saturn's middle atmosphere discussed in Section \ref{discuss}. 

\section{Polar Observations}
\label{obs} 

\subsection{Cassini/CIRS Data}
Coverage of Saturn's high latitudes has been provided by Cassini in clusters of observations whenever the orbital inclination rose out of Saturn's equatorial plane.  Our first study of Saturn's polar region \citep{08fletcher_poles} used data acquired during a `$180^\circ$ transfer' manoeuvre between July 2006 and June 2007, when the orbital inclination was increased via multiple Titan flybys.  Since then, Cassini has returned to these inclined orbits on two further occasions.  Cassini ended its prime mission (2008) and started its equinox mission in an inclined phase (from September 2007 to April 2009), and is currently in the first inclined phase of the solstice mission (May 2012 to March 2015).  Cassini exceeded an inclination of $40^\circ$ from October 2012 to October 2014, and reached a peak inclination of $61.7^\circ$ in April-May 2013.  A second and final inclined phase is planned for  January to November 2016, before the start of Cassini's proximal orbits (close-in, high-inclination phase) at the end of the solstice mission.

Saturn's polar temperatures and gaseous composition can be determined by inversion of mid-infrared spectra from CIRS. This study utilises only one of the two interferometers that comprise the CIRS instrument, focusing on the 7-16 $\mu$m (600-1400 cm$^{-1}$) spectra measured by the Michelson interferometer, which has a higher spatial resolution than observations taken using the longer wavelength ($\lambda>16 \mu$m) polarising interferometer.  Spectra are recorded with two $1\times10$ HgCdTe focal plane arrays (FP3, 600-1100 cm$^{-1}$; and FP4, 1100-1500 cm$^{-1}$) with an instantaneous field of view of $0.27\times0.27$ mrad yielding spatial resolutions of 0.5-3.0$^\circ$ latitude depending on the type of observation.  Typically, the CIRS team designs a range of observations that are given priority on the spacecraft for a set duration (multiple hours, usually covering one or two Saturn rotations).  One of the key advantages of CIRS is its tuneable spectral resolution from 0.5 to 15.0 cm$^{-1}$ (apodised): it is usual for the lowest spectral resolution to be used in scanning the focal planes from north to south along Saturn's central meridian as it rotates, whereas the higher spectral resolutions are used in a `nadir stare mode' at one latitude as the planet rotates beneath (with the arrays oriented to provide some north-south coverage, typically over approximately 10$^\circ$ of latitude).  Previous studies of Saturn's thermal structure have solely focused on these observations for which CIRS was given priority on the spacecraft \citep[e.g., 15-cm$^{-1}$ data has been extensively used by ][]{10fletcher_seasons, 13sinclair}.  However, CIRS also `rides along' with the other remote sensing instruments to acquire non-targeted spectra, primarily at 2.5 cm$^{-1}$ resolution.  For Saturn's high latitudes, these prove to be an invaluable resource, provided that careful filtering is used to remove corrupted spectra.

\begin{figure*}[htbp]
\begin{center}
\includegraphics[width=16cm]{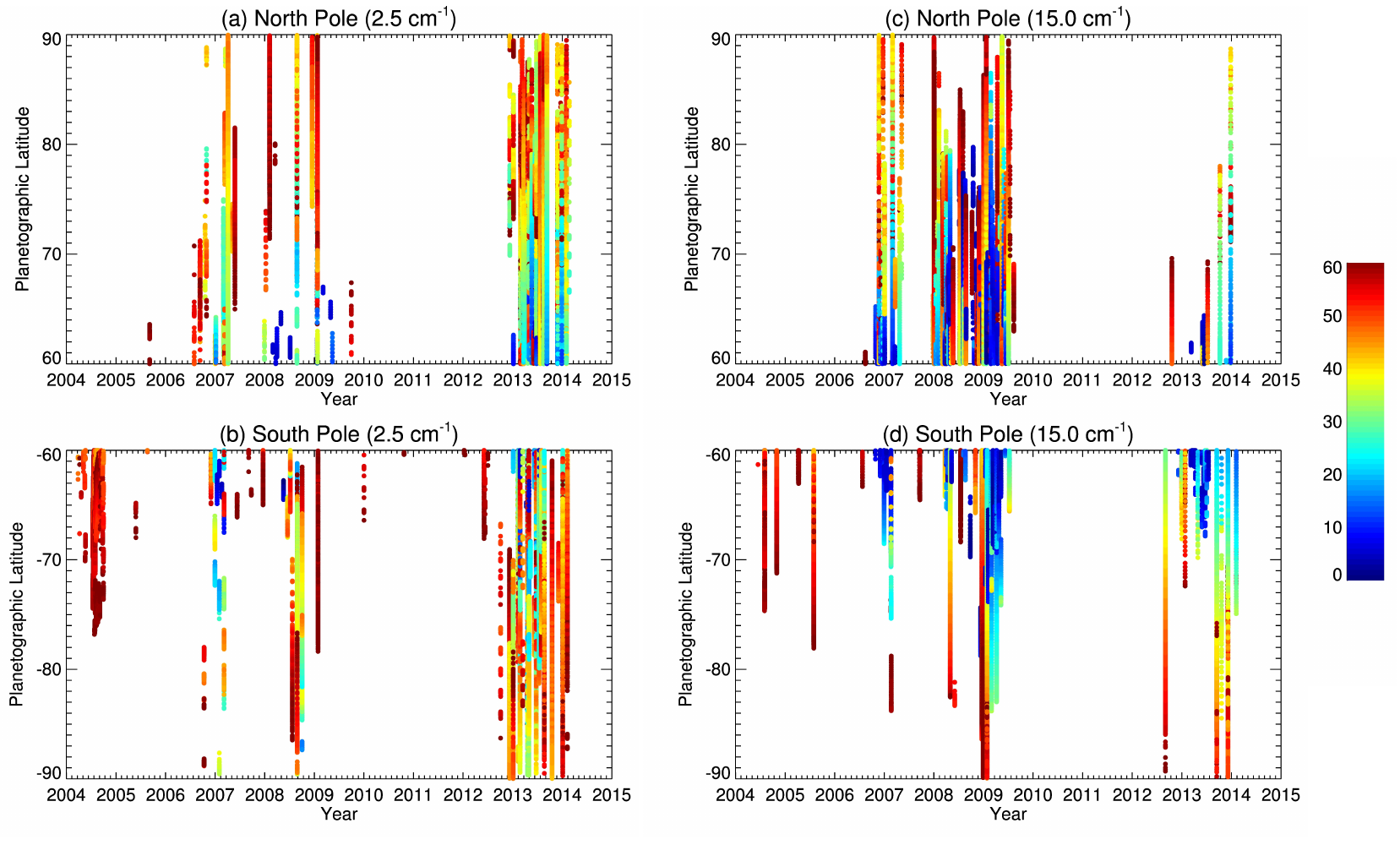}
\caption{CIRS coverage of the north and south poles at spectral resolutions of 2.5 cm$^{-1}$ and 15.0 cm$^{-1}$ on the left and right, respectively.  Each point represents the latitude of the CIRS pixel footprint, binned on a monthly grid.  The colours represent zenith angles from zero to 60$^\circ$, the maximum considered in this study (see the scale bar on the right hand side).  The increased density of observations since 2012 is due to the use of `ride-along' observations during the recent high-inclination phases.}
\label{coverage}
\end{center}
\end{figure*}

Fig. \ref{coverage} shows the polar coverage of the 2.5-cm$^{-1}$ and 15.0-cm$^{-1}$ datasets (both targeted and non-targeted spectra) over a decade of Cassini observations.  The sparse sampling before 2007 and from 2009-2012 is because Cassini was in a near-equatorial orbit during those periods.  The observations have been colour-coded by zenith angle to indicate where the information is coming from - higher zenith angles probe higher altitudes, whereas lower zenith angles are optimal for a combined study of both tropospheric and stratospheric structure.  As the 15-cm$^{-1}$ data fill in some of the gaps in the 2.5-cm$^{-1}$ data, we have elected to use both resolutions in this study.  Spectra have been binned on a 1-month time grid and a 1$^\circ$ latitude grid (in 2$^\circ$-wide latitude bins).  Although coverage of a full latitude circle is impossible for each date, we take the coadded spectra as representative of the zonal mean.  Coadded spectra have been filtered to omit spectra with (i) the CIRS field of view moving rapidly across the target; (ii) significant negative calibrated radiances; (iii) zenith angles exceeding 60$^\circ$; (iv) the full focal plane not entirely on Saturn's disc; and (v) fewer than ten observations contributing to the coadd.  For each month and latitude, spectra were coadded within $\pm10^\circ$ of the mean zenith angle to avoid artefacts from combining data over too broad a range.  

Data were obtained from the latest CIRS calibration pipeline, modified to use 4000 deep space reference spectra for every calibrated target spectrum to improve the signal-to-noise ratio of the observations.  Radiometric random uncertainty is wavelength-dependent (i.e., depending on the NESR, the noise-equivalent spectral radiance of the detectors) and depends on the number of spectra in each coadd. Following \citet{06teanby_halides}, we estimate measurement uncertainties conservatively by taking the larger of (i) the standard error on the mean coadded spectrum or (ii) the expected standard error based on the NESR, reduced by the number of target and calibration spectra.

\begin{figure*}[htbp]
\begin{center}
\includegraphics[width=16cm]{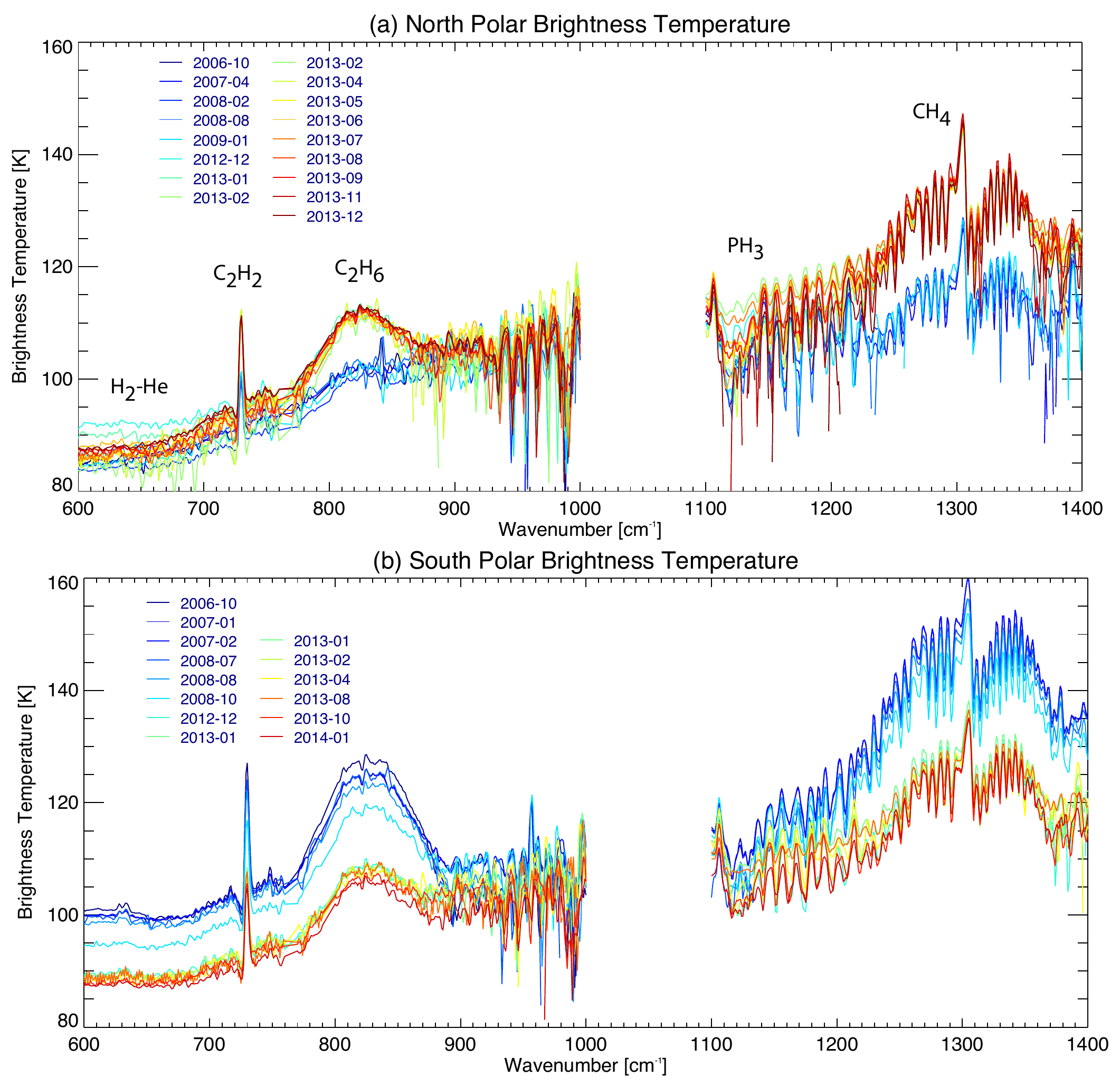}
\caption{Brightness-temperature spectra for the north and south pole showing the evolution over approximately 8 years.  Blue colours represent spectra early in the mission; red colours represent those acquired late in the mission.  Although the brightness temperature depends on both the physical temperature and the zenith angle, this can be used as a reasonable proxy for atmospheric temperature at the peak of the contribution function for each wavenumber.  Spectra were averaged over the whole month within 5$^\circ$ of the pole and within 10$^\circ$ of the mean emission angle for the data set.  Key emission bands are labelled in the top panel.  Radiometric uncertainties vary from observation to observation due to the changing number of spectra in each coadd, but we estimate 0.1-0.3 K near 600-700 cm$^{-1}$, rising to 0.2-0.6 K in the vicinity of the C$_2$H$_2$ and C$_2$H$_6$ features and 0.5-1.5 K near PH$_3$ absorption features.  Uncertainties near the CH$_4$ emission features range from 0.1-0.8 K.  }
\label{pole_spx}
\end{center}
\end{figure*}

Representative spectra from within $\pm5^\circ$ latitude of each pole are shown in Fig. \ref{pole_spx}, highlighting how the spectral shape has altered over the duration of the Cassini mission.  The changing brightness temperatures in the 600-700 cm$^{-1}$ and 1200-1400 cm$^{-1}$ regions reveal varying tropospheric and stratospheric temperatures, respectively, whereas the changing size of the C$_2$H$_2$ $\nu_5$ and C$_2$H$_6$ $\nu_9$ bands between 700-900 cm$^{-1}$ are due to variations in both temperatures and hydrocarbon abundances.  Indeed, the north polar temperatures were so low in the earliest part of the mission that the hydrocarbon emissions are barely discernible in the spectra.  Regions between 1000-1100 cm$^{-1}$ are dominated by PH$_3$ absorption, but the absence of a temperature constraint at the peak of the PH$_3$ altitude sensitivity (near 500-700 mbar) means that we therefore have no corresponding constraints on PH$_3$ variability in this study.  Finally, evidence for methyl acetylene $\nu_9$ and diacetylene $\nu_8$ emission is visible in the south polar spectra near 630 cm$^{-1}$ in the earliest phase of the mission when the temperatures were highest, but is not yet apparent in the north polar spectra.  Coadded spectra of this quality will be analysed in Section \ref{method}, but in the next subsection we discuss the spatial variation of brightness temperature evident in the CIRS maps.

\subsection{Polar Brightness Temperature Maps}

\begin{figure*}[htbp]
\begin{center}
\includegraphics[width=18cm]{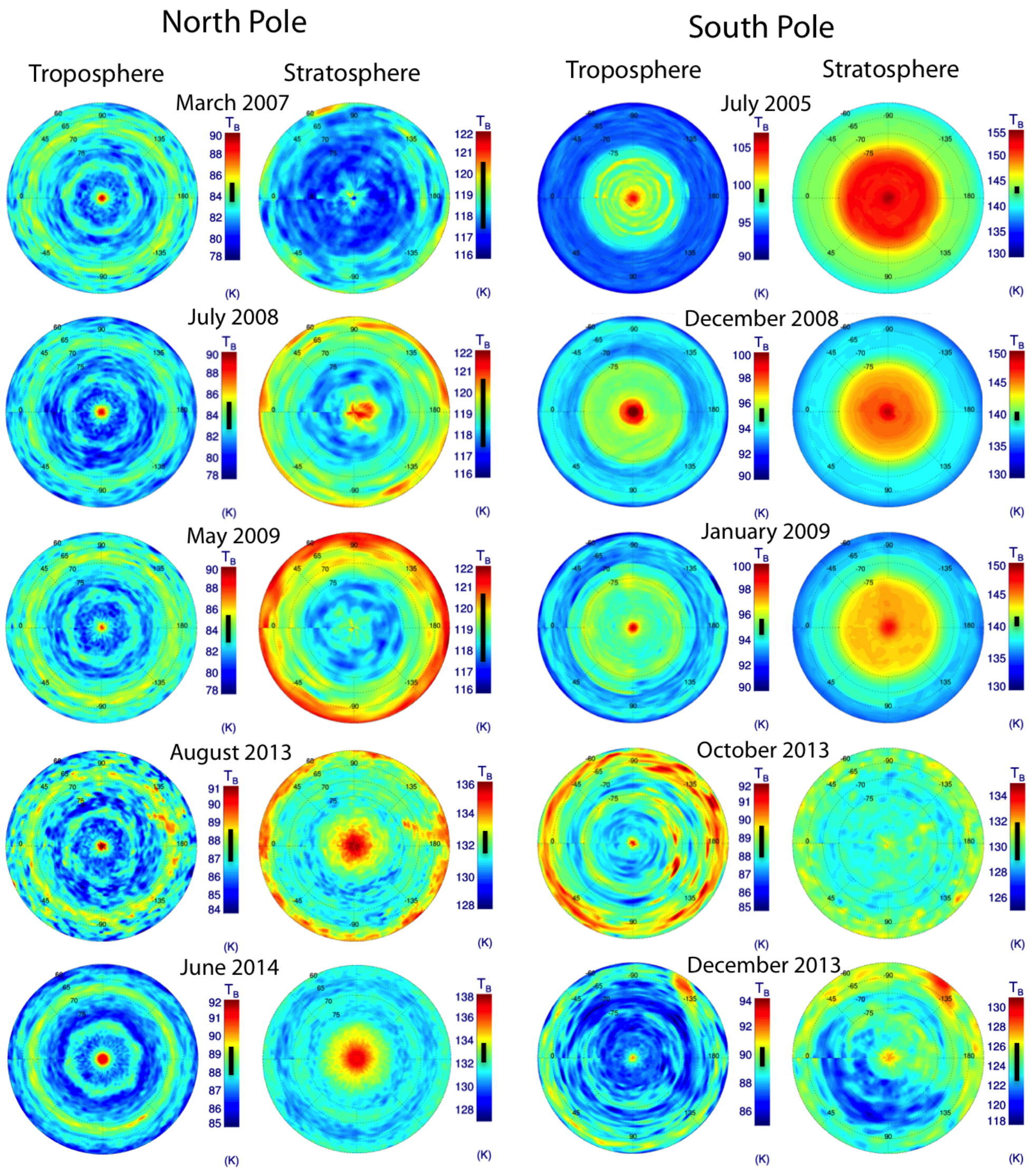}
\caption{Orthographic projections of brightness temperatures to show the approximate distribution of emission at the north (left) and south (right) poles.   Tropospheric maps use the H$_2$-He emission between 600-620 cm$^{-1}$ to sense the 150-mbar level; stratospheric maps use methane emission between 1290-1310 cm$^{-1}$ to sense the 1-mbar level.  The colour scale varies from image to image to maximise the visibility of thermal contrasts, but the uncertainty on the interpolated data is shown by the black bar overlaid on each scale bar (smallest for the hot southern stratosphere, largest for the cold northern stratosphere).  The absolute values of $T_B$ depend both on the physical temperature and the observation zenith angle, which can vary across the map, and this is particularly true of the map in December 2013. Triangular interpolation, in addition to longitudinal smoothing that varied as a function of latitude, has been used to reduce pixel-to-pixel variance in each map.}
\label{polarmap}
\end{center}
\end{figure*}

Polar brightness temperature maps (Fig. \ref{polarmap}) can be assembled from CIRS observations when the spatial coverage is sufficiently dense in a particular month.  The absolute values of the brightness temperatures also depend upon the viewing geometry, due to tropospheric limb darkening and stratospheric limb brightening.  However, these maps are useful to show the morphology of the emission.  We average spectra over 600-620 cm$^{-1}$ (H$_2$ and He collision-induced continuum emission, sensing 150 mbar) and 1290-1320 cm$^{-1}$ (methane emission, sensing 1 mbar) using both 2.5- and 15.0-cm$^{-1}$ resolution data, and show examples of the maps with the best spatial coverage and noise characteristics between 2004 and 2014 in Fig. \ref{polarmap}.   The random uncertainty on the maps (shown by black bars in each scale bar) varies with time, with the cold north pole being noisiest at the start of the mission when the brightness temperatures were lowest.  

The northern hexagon can been seen as a persistent warm belt before and after Saturn's 2009 equinox, when it became visible to Cassini's visible-light cameras and ground-based observers \citep{14sanchez}.  They calculated a mean planetographic latitude for the hexagon vertices of $77.5\pm0.2$$^\circ$: the warm band in the CIRS images sits on the poleward side of this prograde jet such that $dT/dy$ is positive across the jet \citep[the zonal jet weakens with altitude according to the thermal windshear relation, as shown by][]{08fletcher_poles}.  Furthermore, \citet{14sanchez} demonstrated that the hexagon was moving relative to the System III longitude system \citep[defined by the Voyager-era kilometric radiation measurements, e.g.,][]{07seidelmann} by $0.0128\pm0.0013^\circ$ per day, equivalent to a translation of $\approx30^\circ$ westward between March 2007 and August 2013.  This is consistent with the rotation observed in Fig. \ref{polarmap}, although the sparse temporal sampling and pixel-to-pixel noise prevents an accurate determination of the drift rate from the CIRS data alone.


The cyclonic north polar vortex (NPC) is always present throughout the tropospheric observations, but the structure in the stratosphere is dominated by noise until the most recent (2013) observations.  Subtle warming is seen towards the north pole before the equinox, but the August 2013 images show a dramatic increase in methane emission within 5$^\circ$ of the north pole.  This stratospheric vortex has yet to expand to the same dimensions as that observed at the south pole (i.e., extending $15^\circ$ from the pole), but the pole is clearly warming at a faster rate than the lower latitudes.

The south polar region has changed dramatically in both the troposphere and stratosphere as the pole has moved into the darkness of autumn.  The SPC remains visible in both the troposphere and stratosphere within a few degrees of the pole, but the atmosphere surrounding the SPC has cooled considerably.  In the troposphere, the poleward-increasing temperatures of 2005 were replaced by a distinct cool polar zone by 2013, so that the warm polar belt centred on 75$^\circ$S (i.e., on the poleward side of the prograde jet at 74$^\circ$S) is now readily discernible in the images.  Although both hemispheres feature warm tropospheric polar belts between 75-80$^\circ$, these images show that it is more poleward in the north than in the south.  In the stratosphere, the warm `hood' has almost entirely dissipated, leaving a south polar stratosphere resembling the cold north pole in the earliest stages of the Cassini mission.   As complete polar coverage is limited to a small handful of dates, the subsequent analysis reported in this study will deal with zonal mean spectra to construct 2D cross-sections (pressure and latitude) through the polar regions.

\section{Spectral Inversion}
\label{method}

CIRS zonal mean spectra for each latitude and month are inverted using an optimal estimation \citep{00rodgers} spectral-retrieval algorithm \citep[NEMESIS,][]{08irwin}, using a Levenburg-Marquardt iterative scheme to minimise a two-term cost function comprised of the residual fit to the data (weighted by the measurement error covariance matrix) and our \textit{prior} knowledge of the atmospheric state vector (weighted by the \textit{a priori} uncertainty covariance matrix).  The latter term ensures smooth and physically-realistic retrieved state vectors.  Random uncertainties on retrieved variables are provided by the \textit{posteriori} error matrix \citep{08irwin}, and any potential systematic bias towards \textit{a priori} assumptions is held constant by using the same prior for all inversions.  Sources of spectral line data are identical to those presented in Table 4 of \citet{12fletcher}, and are used to pre-tabulate $k$-distributions (ranking absorption coefficients from high-resolution spectral line databases according to their frequency distributions) with a spectral resolution appropriate for the CIRS 2.5 cm$^{-1}$ and 15.0 cm$^{-1}$ spectra and a triangular instrument function (an approximation to the Hamming apodisation).  



The \textit{a priori} $T(p)$ (a low-latitude mean of nadir and limb retrievals) is defined on 120 pressure levels equally spaced in $\log(p)$ between 1 $\mu$bar and 10 bar.  Retrieved temperatures will relax back to these \textit{a priori} profiles when the information content of the spectrum is low, such that temperatures outside of the 75-250 mbar region (troposphere from H$_2$-He collision-induced absorption in FP3) and the 0.5-5.0 mbar region (stratosphere from methane emission in FP4) are simply estimated by smooth interpolation between the retrieved profile and the \textit{a priori}.  Based on retrieval tests by \citet{10fletcher_seasons}, we find that a simultaneous retrieval of temperature and composition provides the best fit to the nadir data (i.e., rather than retrieving temperature first, followed by molecular abundance in a two-step process). This is due to the slightly different altitude sensitivities for the temperature and compositional sounding.  Indeed, by performing this simultaneous retrieval, we have no need to vary the vertical profiles of the hydrocarbons with altitude.  Instead, we simply scale global-mean hydrocarbon profiles provided by CIRS limb spectroscopic analysis \citep{09guerlet, 10guerlet}, and report the molecular abundance near the pressure of peak sensitivity.  The CIRS limb profiles provided by Guerlet et al. were themselves based on \textit{a priori} information from the 1-D diffusive photochemistry models of \citet{05moses_sat}, which show that CIRS nadir spectra probe much lower altitudes than the hydrocarbon source regions at microbar pressures.  Although the resulting vertical profiles will be inaccurate (the vertical gradients are likely changing with time), we find that this has no effect on our ability to reproduce the CIRS spectra at 2.5 cm$^{-1}$ and 15.0 cm$^{-1}$ spectral resolution, and this choice of a single scaling factor allows us to treat all spectra in a consistent fashion.

We assume that Saturn's stratospheric methane abundance is spatially and temporally uniform at a mole fraction of $4.7\times10^{-3}$ \citep{09fletcher_ch4}.  The methane abundance decreases with altitude due to photochemistry and diffusive processes \citep{00moses}, and the homopause height is known to be variable at microbar pressures \citep[e.g.,][]{09nagy}.  The nadir spectra considered here do not probe such low pressures, and \citet{05moses_sat} predict no seasonal variations in methane's abundance at the millibar pressures probed in this study. The inversion assumes an aerosol-free atmosphere, consistent with the predicted low opacity of Saturn's tropospheric hazes observed at near-infrared wavelengths \citep[e.g.,][]{09west}.  Our \textit{a priori} PH$_3$ profile (a well-mixed abundance for $p>550$ mbar, decreasing with height due to photochemistry at $p<550$ mbar) is based on the global mean results of \citet{09fletcher_ph3}.  Rather than retrieving a parameterised profile (e.g., varying the deep abundance and vertical variation), we avoid the degeneracy between these parameters by simply scaling this profile during the retrieval of temperatures and hydrocarbon abundances to improve the fit to the 1100-1200 cm$^{-1}$ region.  This permits consistent treatment of all CIRS spectra at the two different resolution settings.  Ammonia is considered to be spatially homogeneous, given the low sensitivity of the CIRS mid-IR spectra to this molecule \citep{12hurley}.  


\section{Results}
\label{results}

CIRS 600-1400 cm$^{-1}$ spectra with spectral resolutions of 2.5 cm$^{-1}$ and 15 cm$^{-1}$ were initially treated separately on a month-by-month basis.  We simultaneously retrieve a vertical $T(p)$ profile and scale factors for the \textit{a priori} profiles of phosphine, acetylene and ethane.  We used FP3 spectra from 600-1000 cm$^{-1}$ (omitting a region near 765 cm$^{-1}$ known to be contaminated by an electrical spike) and FP4 spectra from 1100-1370 cm$^{-1}$.  Note that we omit wavenumbers longer than 1370 cm$^{-1}$, which feature stratospheric emission from C$_3$H$_8$ (propane at 1376 cm$^{-1}$) and C$_2$H$_6$ ($\nu_6$ at 1379 cm$^{-1}$) that are not included in our line database.  Retrievals were tested for sensitivity to initial conditions, and we found that we had to decrease the measurement uncertainty near to the CH$_4$ emission to ensure optimal fitting of the stratospheric temperatures (i.e., a stronger weighting to the data than to the \textit{a priori} when minimising the cost function).  Once the retrievals were complete, the 2.5-cm$^{-1}$ and 15.0-cm$^{-1}$ resolution results were compared and combined to create a 3D matrix of state vector parameter (either temperature or composition), latitude and time.  

\subsection{Zonal mean polar temperatures}
\label{Tres}

\begin{figure*}[htbp]
\begin{center}
\includegraphics[width=10cm]{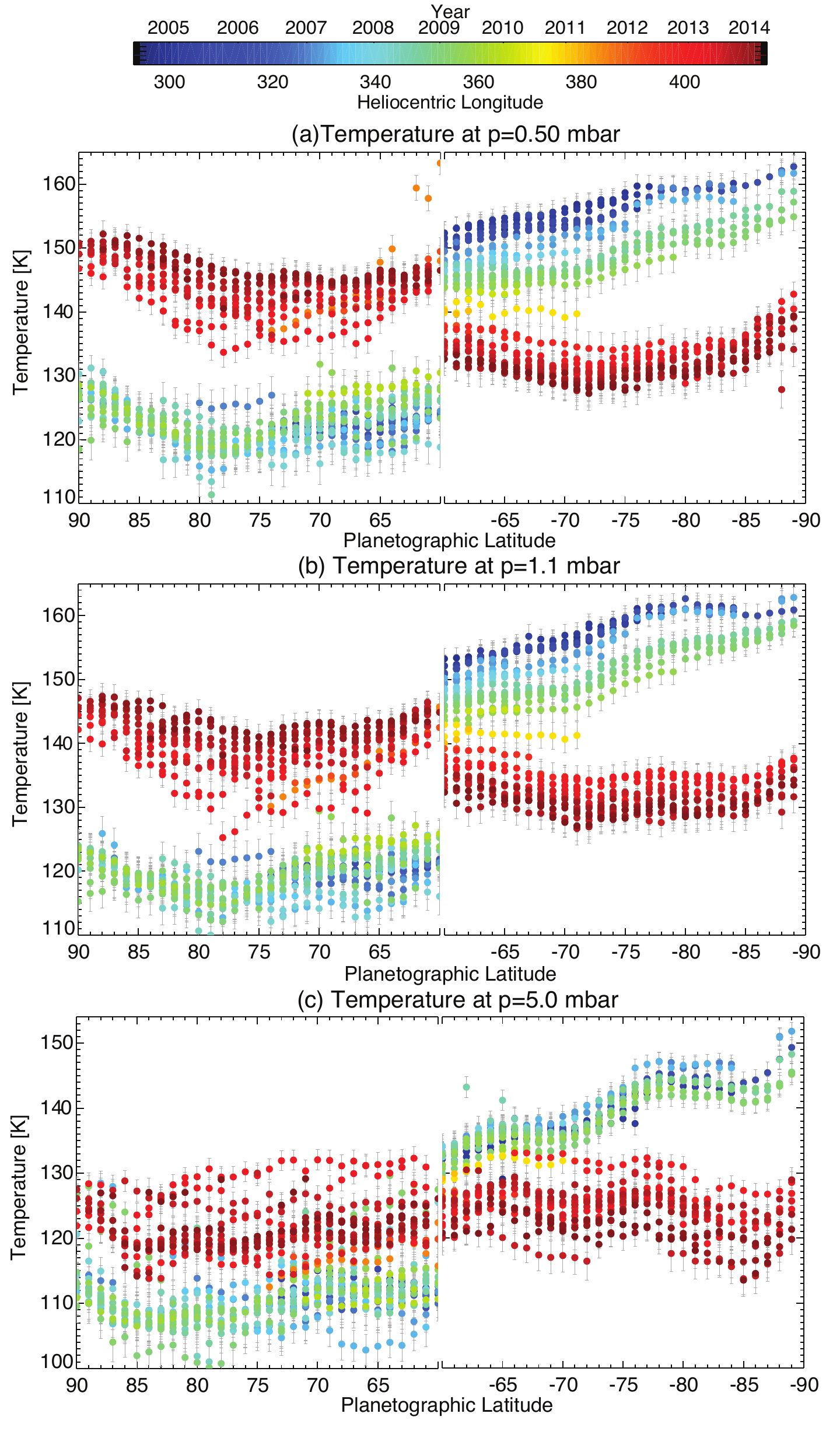}
\caption{Zonal mean polar temperatures as a function of latitude and time at three representative atmospheric levels in the stratosphere.  The colour of each point refers to the dates and planetocentric solar longitudes in the key at top of the figure.  The gap in polar coverage between 2009 and 2012 is evident in these figures. The uncertainties on the individual retrievals are (on average) 2.3-2.6 K at 0.5 mbar; 2.2-2.4 K at 1 mbar; and 2.3-2.5 K at 5 mbar.}
\label{Tlat_strat}
\end{center}
\end{figure*}

\begin{figure*}[htbp]
\begin{center}
\includegraphics[width=10cm]{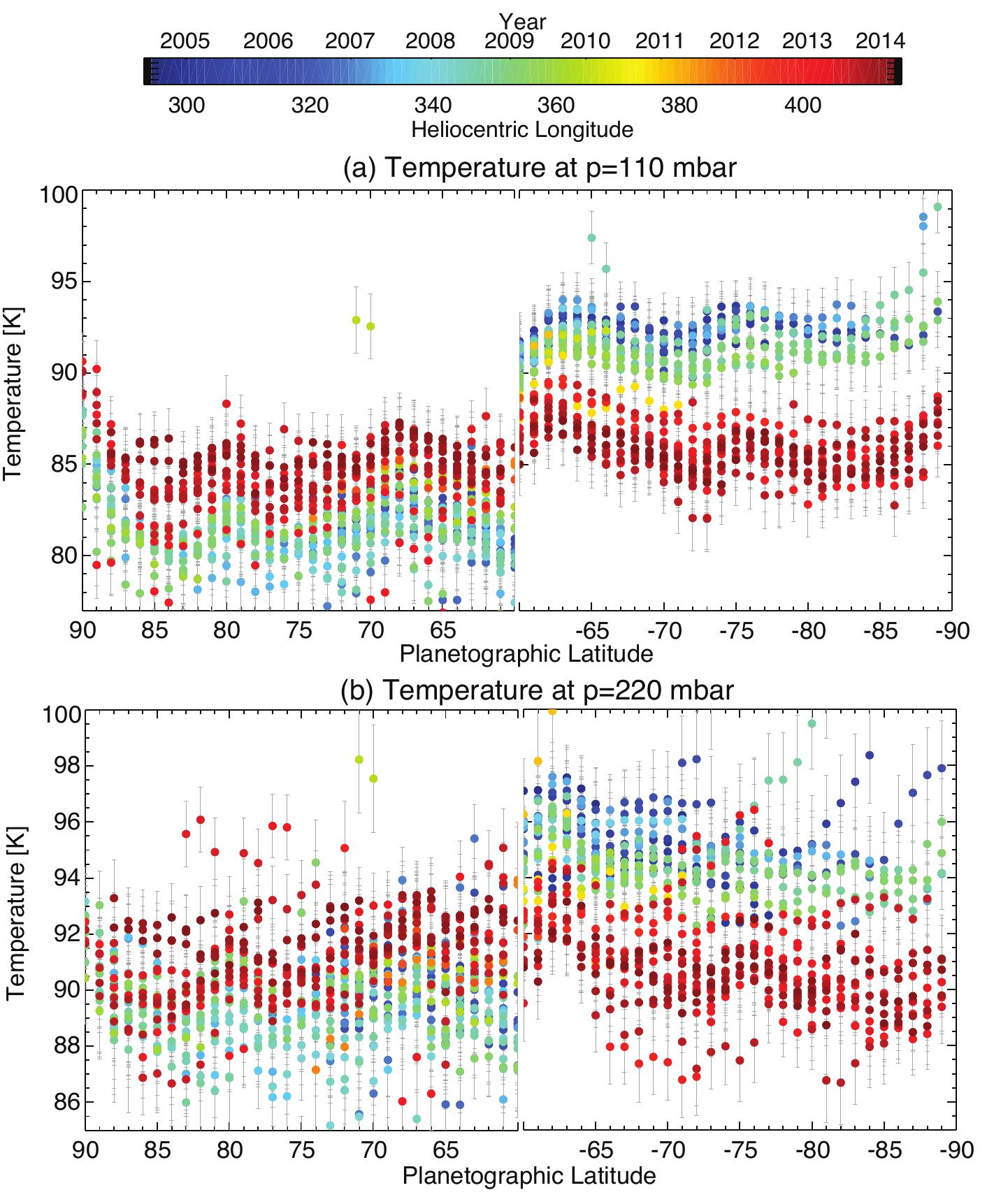}
\caption{Zonal mean polar temperatures as a function of latitude and time at two representative atmospheric levels in the upper troposphere.  The colour of each point refers to the dates and planetocentric solar longitudes in the key at top of the figure.   The uncertainties on the individual retrievals are (on average) 1.6-1.7 K at 110 mbar and 1.3-1.5 K at 220 mbar.}
\label{Tlat_trop}
\end{center}
\end{figure*}

Figs. \ref{Tlat_strat} and \ref{Tlat_trop} present zonal mean temperatures as a function of latitude and time for five representative tropospheric and stratospheric pressure levels.  The south polar cooling and north polar warming is clearly seen for all $p<220$ mbar, but higher pressures do not show any significant variations over time.  The tropospheric spikes in temperatures associated with the NPC and SPC are evident in the upper troposphere and south polar stratosphere, but the north polar stratospheric warming does not show such a strong latitudinal gradient. The gap in our polar time coverage between 2009 and 2012 is evident in these figures, where the retrieved temperatures fall into two distinct groups (2005-2008 and 2013-2014). The magnitudes of the seasonal temperature changes will be discussed in Section \ref{compare} and compared to the expectations of simple radiative models.  Furthermore, the changing latitudinal temperature gradients have implications for the vertical shears on the zonal winds, as discussed in Section \ref{winds}.  


The CIRS latitude sampling is far from consistent from month to month, and Figs. \ref{Tlat_strat}-\ref{Tlat_trop} reveals numerous outliers that appear to be spurious.  Accurate calibration of an interferogram requires that reference targets (both hot and cold) be observed under near-identical conditions (instrument temperatures, voltages, etc.) as the science target, which is rarely the case.  Although sophisticated calibration techniques are employed to minimise these instrumental differences, spectra with baseline offsets are sometimes produced, which are then interpreted literally by spectral inversion algorithms and result in temperature and composition measurements that deviate from the general trends.  Unfortunately, these spurious measurements cannot be rejected based on the quality of the spectral fit, so we choose to include them but use smooth interpolation through the temporal trends to minimise their influence.  We therefore `reconstruct' the temperature field by fitting a smooth quadratic function through the sparse data at every latitude and pressure level using a robust non-linear curve fitting technique in IDL \citep[MPFIT,][]{09mpfit}.   The regression for each pressure level was weighted by the uncertainties on the temperatures at each latitude and date (e.g., those in Figs. \ref{Tlat_strat}-\ref{Tlat_trop}).  The purpose of this reconstruction is to interpolate through sparsely-sampled measurements whilst minimising the influence of outliers.  

\begin{figure*}[htbp]
\begin{center}
\includegraphics[width=17cm]{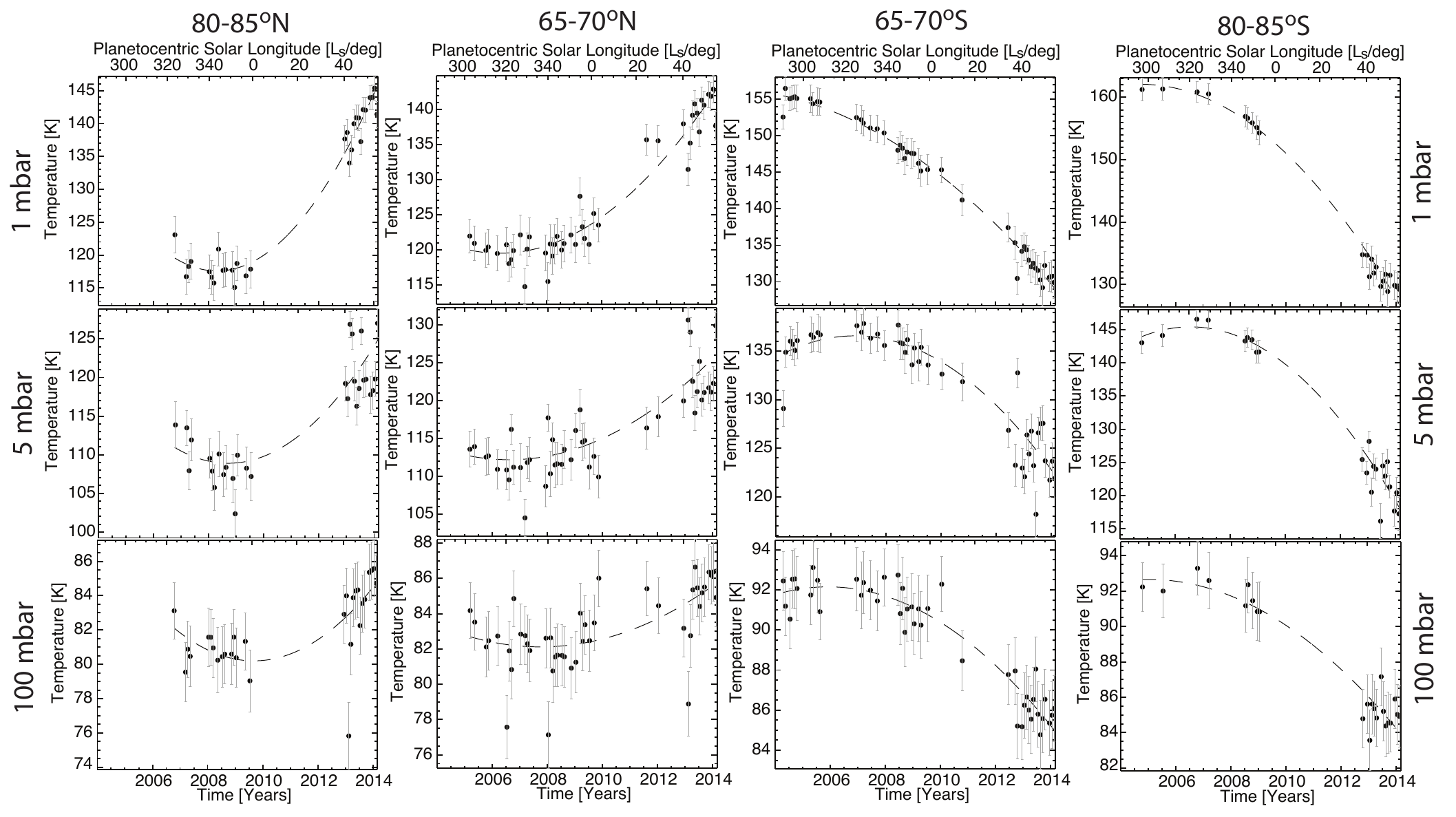}
\caption{Variation of temperatures with time and plantocentric solar longitude ($L_s$) at 1, 5 and 100 mbar for four representative (but arbitrary) latitudes in the north and south polar regions.  Uncertainties on the measurements (shown as grey bars) are estimated as the mean of the formal retrieval errors shown in Figs. \ref{Tlat_strat}-\ref{Tlat_trop} over this $5^\circ$ latitude range on the monthly grid.  Some outliers are evident due to systematic uncertainties in the data.  Smooth quadratic functions are fitted through these sparse data to reconstruct the temperature variations, although the finer latitude grid ($1^\circ$) was used in subsequent analysis.  }
\label{Tchange}
\end{center}
\end{figure*}

\begin{figure*}[htbp]
\begin{center}
\includegraphics[width=18cm]{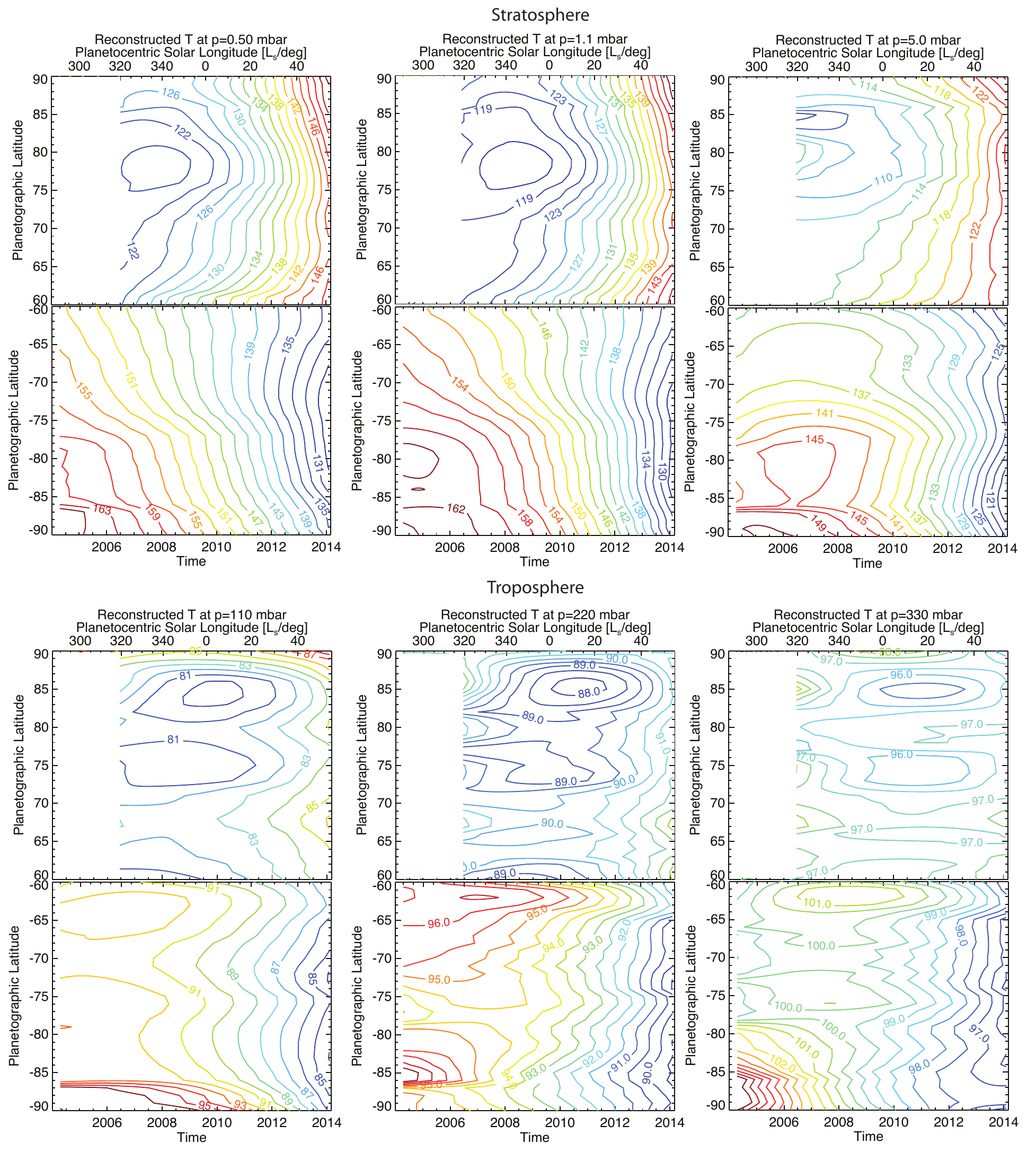}
\caption{Temperature variability at six representative stratospheric (upper three panels) and tropospheric (lower three panels) pressure levels, interpolated from the retrieved temperatures using the quadratic functions introduced in Fig. \ref{Tchange}.  North and south polar temperature changes are compared in each panel. Following Fig. \ref{coverage}, extrapolated northern hemisphere results prior to mid-2006 are omitted.  In the stratosphere, the 2-K contour spacing is equivalent to the temperature retrieval uncertainties (2.3-2.6 K at 0.5 mbar; 2.2-2.4 K at 1 mbar;  2.3-2.5 K at 5 mbar).  In the troposphere, the contour spacing is smaller than the uncertainties to highlight trends (1.6-1.7 K at 110 mbar; 1.3-1.5 K at 220 mbar; 1.4-1.6 K at 330 mbar).}
\label{Trecons}
\end{center}
\end{figure*}

The retrieved temperatures at three pressure levels (1, 5 and 100 mbar) are shown for four latitudes in Fig. \ref{Tchange}, with the quadratic function providing the best-fit reconstruction of the temperature trends with time.  Although a skewed-sinusoid might be more appropriate over longer time periods (skewed because heating and cooling rates are not necessarily equivalent either side of the temperature maxima), the quadratic fits were found to be acceptable over the ten-year span of these data.  The fits also reveal where temperature maxima and minima have occurred within the span of the CIRS observations, even if the time when $\pderiv{T}{t}=0$ was not observed directly in the sparse dataset. The lag between solstice and the temperature turning points (minima in the north, maxima in the south) generally increases with depth, consistent with the longer radiative timescales in the deeper atmosphere.  Intriguingly, the 5-mbar lag in the southern stratosphere appears to be greater than the lag at 100 mbar, suggesting that the atmosphere continues warming up for longer in the mid-stratosphere (i.e., a longer radiative time constant at this altitude than expected due to the presence of aerosols, or the influence of dynamics).  The quadratic fitting process was applied to all retrieved temperatures on a 1$^\circ$ latitude bin, allowing us to create smooth contours of temperature variation over the time span of the observations, shown in Fig. \ref{Trecons}.  It should be noted that these contours represent a smooth idealised case, and that (i) certain epochs of data are missing, particularly the northern latitudes in 2004-2006 (as shown in Fig. \ref{coverage}); (ii) some retrievals result in temperatures that are outliers from this smooth variation, particularly in the troposphere, where Fig. \ref{Tchange} shows a greater standard deviation compared with the stratosphere; and (iii) the smooth functions are for interpolation only, and should not be used to extrapolate to time periods outside of the span of the CIRS observations.


Nevertheless, the reconstructed temperatures in Fig. \ref{Trecons} reveal insights into how Saturn's atmosphere is responding to the changing seasons. Minima in the zonal temperatures were detected in the northern hemisphere, and maxima in the southern hemisphere, during the timespan of the Cassini observations. The timing of these turnover events (where $\pderiv{T}{t}=0$) is later for higher latitudes.  The atmospheric response lags behind insolation changes by approximately a season (the northern winter temperatures are at their lowest near the spring equinox, 6-8 years after winter solstice).  Fig. \ref{Trecons} shows that the northern hemisphere stratospheric minimum is not directly at the north pole, but occurs between 75$^\circ$ to 80$^\circ$N, suggesting that the north polar temperatures are influenced by dynamics as well as radiative balance.  The same is true of the warm tropospheric polar cyclones within 2-3$^\circ$ of each pole, where Fig \ref{Trecons} shows a distinct `kink' in the temperature field and increasing temperatures towards each pole.  The increasing importance of belt-zone temperature contrasts can be seen by comparing the stratospheric temperatures (smoothly varying with latitude) with the tropospheric temperatures (showing differing behaviours from warm belts to cooler zones).  The latitudinal contrasts $dT/dy$ will be used in Section \ref{winds} to explore the implications for these thermal changes on the zonal wind shears.  Fig. \ref{Trecons} also indicates a stratospheric boundary near 75$^\circ$S (shown by the distortion of the temperature contours) that encloses the south polar warm hood.  There are no similar boundaries in the northern stratosphere in 2014, even though this region has been warming quickly, suggesting that we are still awaiting the onset of the northern summer vortex observed during the last Saturn year (see Section \ref{intro}). 

\begin{figure*}[htbp]
\begin{center}
\includegraphics[width=12cm]{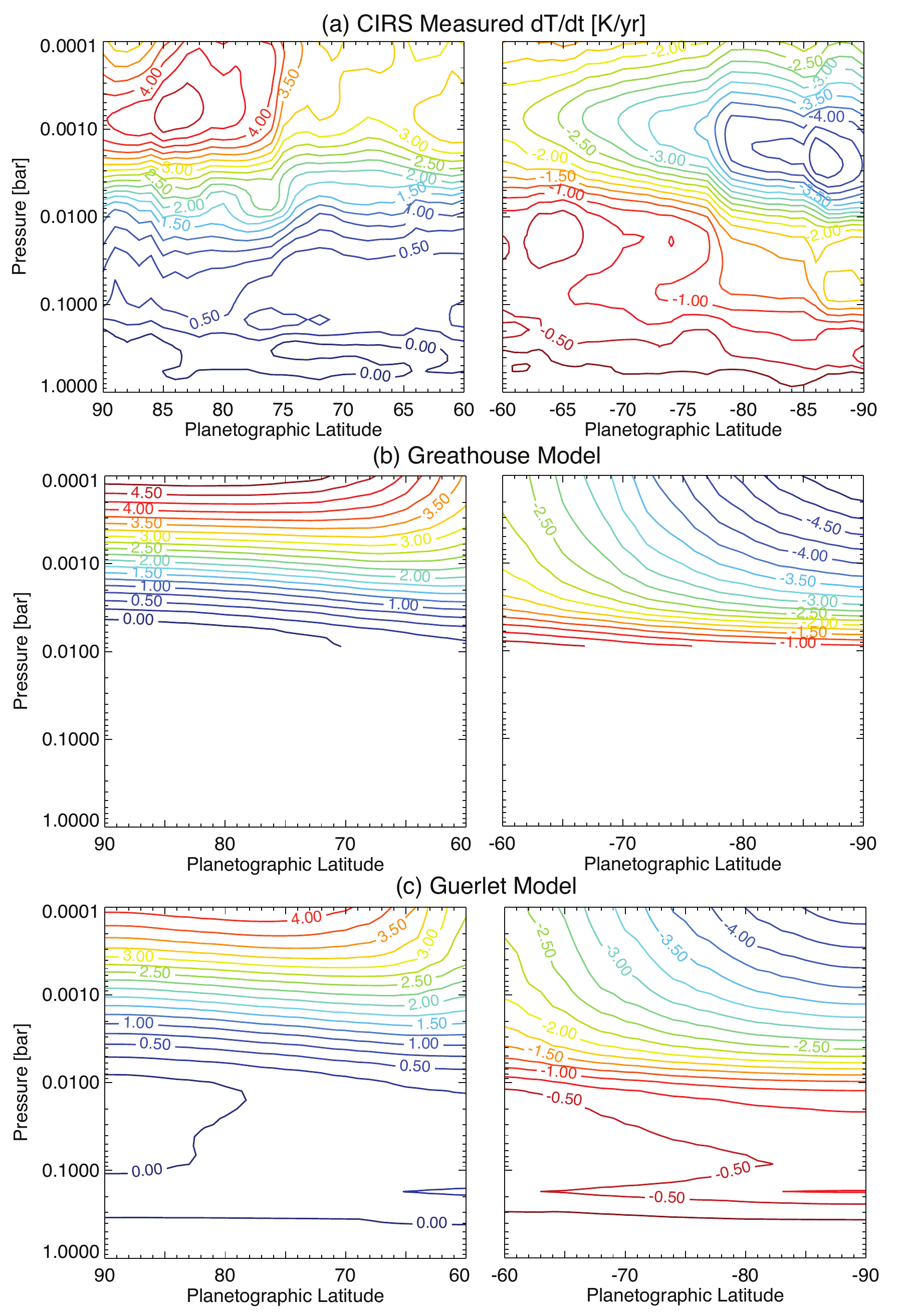}
\caption{(a) Rate of change of temperatures with time over the ten years of observations from a linear fit to the retrieved temperatures (this is an approximation, given that the temperature variations in Fig. \ref{Tchange} are clearly not linear). As such, this figure should be treated as a means of visualising the temperature changes observed by CIRS, and not as an accurate representation of the heating/cooling rates.  Panels (b) and (c) are predicted temperature changes from the radiative equilibrium models of \citet{08greathouse_agu} and \citet{14guerlet}, described in Section \ref{compare}. Note that the Greathouse model is cut off at 10 mbar as it lacks the opacity sources to model the tropospheric variability.  Contours are given for every 0.25 K/yr, which is larger than the uncertainty on the linear $dT/dt$ measurements, namely 0.13-0.14 K/yr near 1-5 mbar region, decreasing to 0.09-0.11 K/yr in the 100-400 mbar region. }
\label{dTdt}
\end{center}
\end{figure*}

Fig. \ref{Trecons} indicates that temperatures have changed with different rates at different locations.  As an aid to visualisation, we fit a simple linear regression to the temperature variations for each latitude and altitude to provide an approximate $\pderiv{T}{t}$ over the ten-year timespan of these measurements, shown in Fig. \ref{dTdt}(a).  This calculation is performed with the strong caveat that $\pderiv{T}{t}$ is clearly not a constant over seasonal timescales (e.g., Fig. \ref{Tchange}), but nevertheless provides a useful graphical representation of the temperature changes with location and depth.  The magnitude of the temperature changes decreases with increasing depth, and the deeper troposphere ($p>400$ mbar) shows no seasonal variability over the course of the Cassini mission.  The rate of warming or cooling is not a simple function of latitude, as one might expect from a simple model with constant abundances of atmospheric heaters/coolers.  Instead, we find the south polar cooling to be most intense within the summer stratospheric hood, and particularly close to the south pole itself.  A similar situation is found in the north polar region, where the most intense warming occurs within 15$^\circ$ of the north pole.  The peak warming in the north is in the 0.5-1.0 mbar region, whereas the peak cooling in the south is in the 1-3 mbar region.  Either localised dynamics (e.g., vertical motions and adiabatic heating/cooling) or entrained hydrocarbons and aerosols (acting as additional contributors to the polar radiative budget) could be responsible for this transition in the heating/cooling rates poleward of 75$^\circ$.  In the next section, we search for variability in the polar composition in an attempt to distinguish between these possibilities.

\subsection{Polar hydrocarbon abundances}
\label{compres}

\begin{figure*}[htbp]
\begin{center}
\includegraphics[width=12cm]{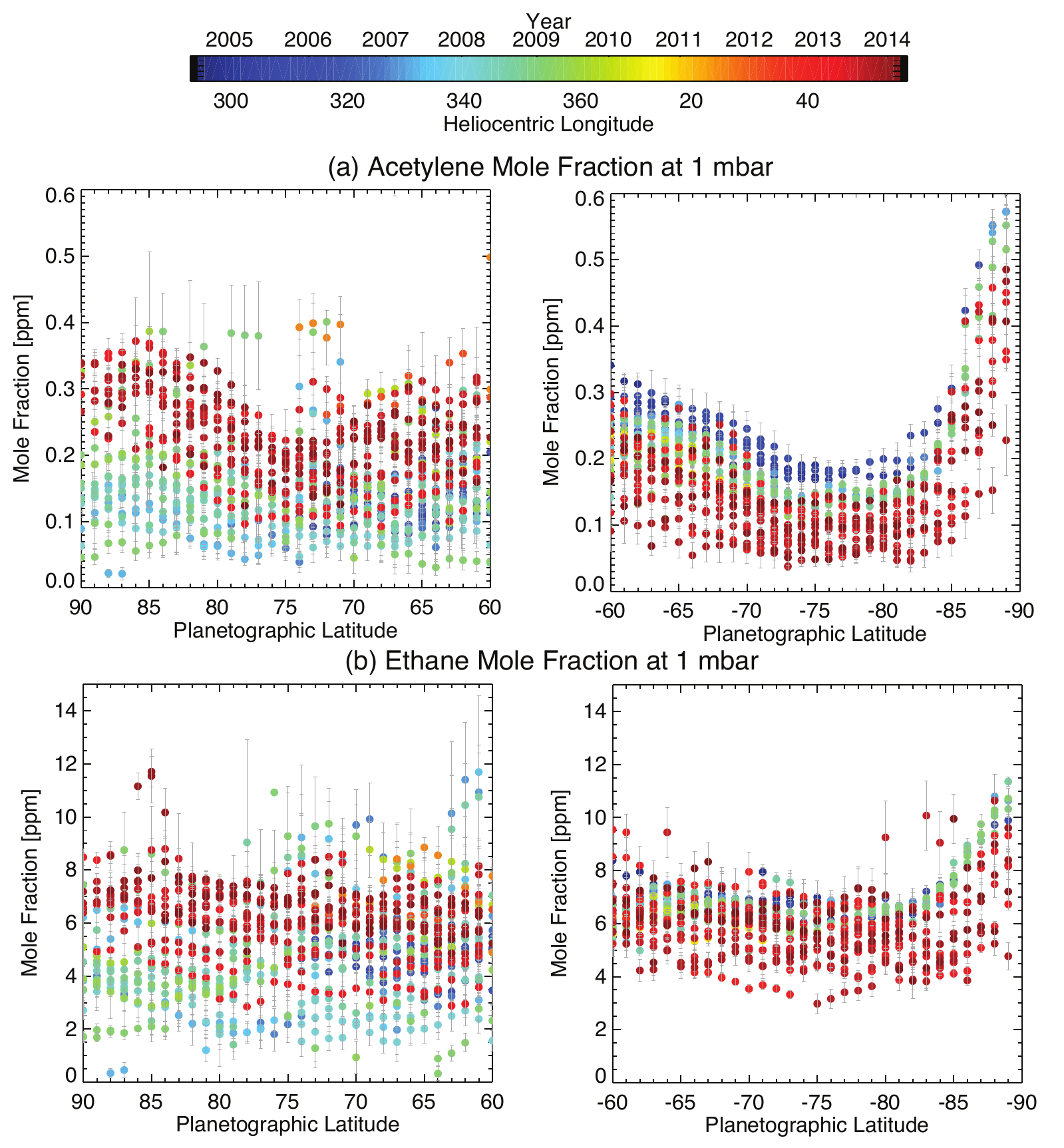}
\caption{Retrieved abundances of stratospheric hydrocarbons at 1 mbar from all CIRS datasets, where the date of the observation corresponds to the legend at the top.  Although the results are highly scattered, the hydrocarbons show some discernible trends with time. }
\label{comp}
\end{center}
\end{figure*}

Unlike atmospheric temperatures, which influence all regions of the CIRS spectra and can therefore be derived over a broad altitude range with small posterior uncertainties, the information on gaseous composition is limited to narrow spectral intervals and is thus more prone to random measurement uncertainties. This is particularly true in the cold northern stratosphere and upper troposphere, where spectral signatures of phosphine, acetylene and ethane are often indistinguishable from the measurement noise.  Rather than coadd spectra on a coarser latitude and temporal grid to increase the signal-to-noise ratio, we instead choose to average retrieved parameters from the fine grid of latitudes and times to understand the compositional variability over the timespan of the Cassini measurements.  This has the side effect that the retrieved scale factors for the \textit{a priori} gas profiles are often extremely noisy, as shown in Fig. \ref{comp}, which shows the mole fractions of acetylene and ethane near the peak of their contribution functions as a function of time.  Stratospheric hydrocarbons show trends with latitude and time, such as the increasing north polar C$_2$H$_2$ abundance at the north pole and the general decreased abundance over the south pole with time (Fig. \ref{comp}a), and the sharp enhancement in stratospheric hydrocarbons within 5$^\circ$ of the south pole that is not evident at the north pole.  We did not detect phosphine variability outside of the scatter of retrieval uncertainty.

\begin{figure*}[htbp]
\begin{center}
\includegraphics[width=17cm]{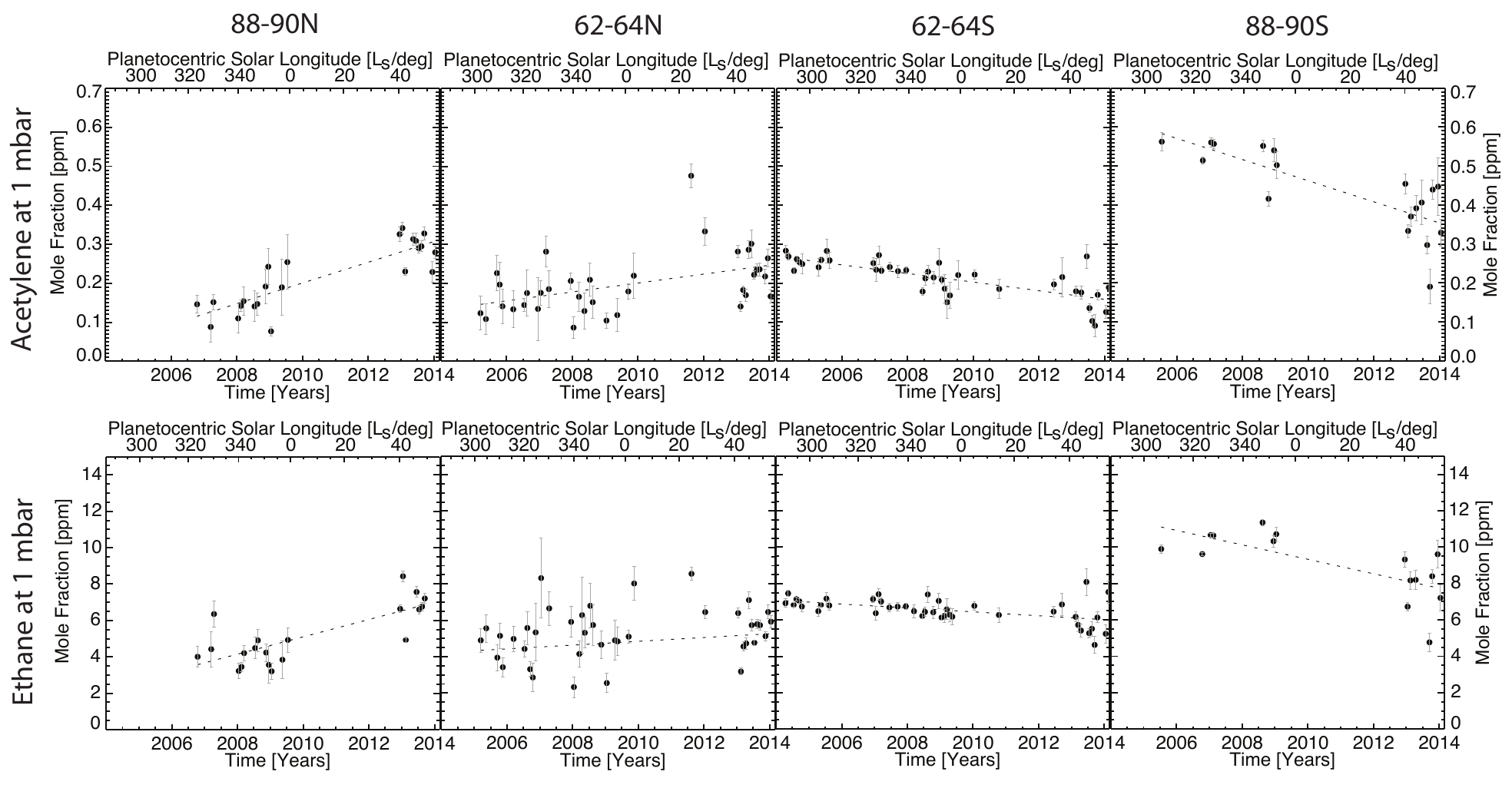}
\caption{Variations of acetylene and ethane at 1 mbar over the 10-year span of observations, shown as a function of time and planetocentric solar longitude for four representative latitudes. Linear regression lines are shown as dotted, which were used to measure the rate of change of the mole fraction at all latitudes in Fig. \ref{dqdt}.  The change in the gradient from 60 to 90$^\circ$ in the hydrocarbon species can be seen in these figures.  The scatter between the individual points is due to lower signal in the retrievals with fewer coadded spectra, and outliers in hydrocarbon abundances are related to the outliers in temperature as discussed in Section \ref{Tres}.   }
\label{compchange}
\end{center}
\end{figure*}

\begin{figure*}[htbp]
\begin{center}
\includegraphics[width=12cm]{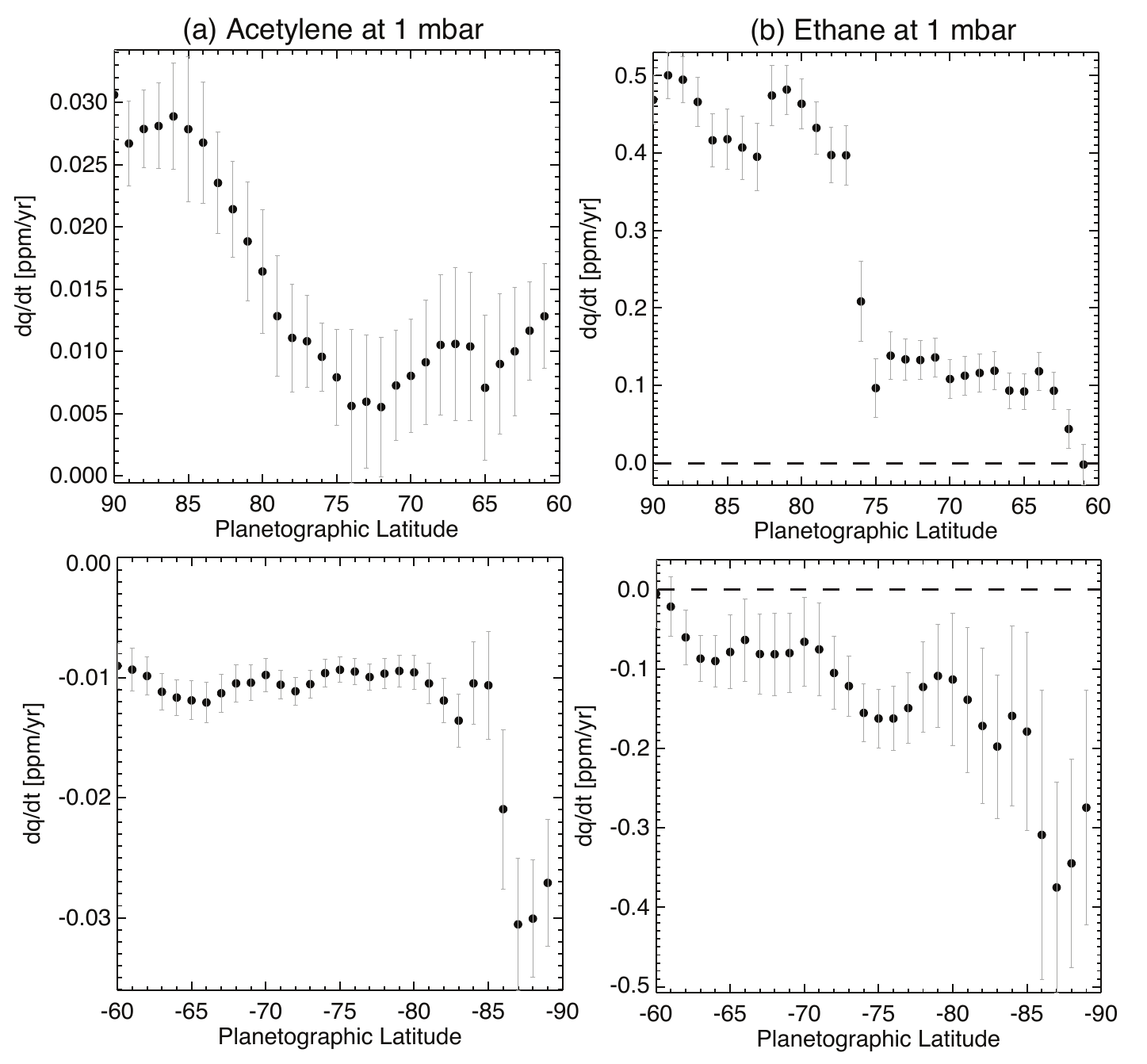}
\caption{Rate of change of 1-mbar composition over the ten-year span of these observations, from a linear fit to the retrieved abundances in Fig. \ref{compchange} on a 1$^\circ$ latitude grid.  North polar variability is shown in the upper row, south polar variability on the lower row.  Larger error bars occur when there are fewer points used in the regression.  The zero point is shown by a horizontal dashed line. }
\label{dqdt}
\end{center}
\end{figure*}

To better understand these temporal trends, we select four representative (but arbitrary) latitudes and plot the hydrocarbon changes with time in Fig. \ref{compchange}.  Unlike the quadratic fits to the temperature trends, we found that linear regression adequately reproduced the observed trends because the scatter of the compositional results was so large.  The regression lines were weighted by the uncertainty in the fits (i.e., the reduced $\chi^2$ calculated for the regions surrounding the gas emission features), which allowed us to reduce the influence of some of the outlying points in Figs. \ref{comp} and \ref{compchange}.  Acetylene and ethane show enhancements in the north and depletions in the south; and it is clear that the gradients are stronger at the highest latitudes than at $\pm60$$^\circ$ latitude.  The outliers for acetylene near 60$^\circ$N in 2011-2012 are due to the dynamic perturbations associated with the storm-produced stratospheric vortex \citep[the `beacon',][]{12fletcher}. These regression fits were performed for all latitudes, and their gradients and associated uncertainties are shown in Fig. \ref{dqdt}, which estimates the rate of change of the mole fraction ($\pderiv{q}{t}$) as a function of latitude over the ten-year span of these data.  Where fewer points were used in the regression (e.g., right at the south pole), the uncertainty in $\pderiv{q}{t}$ is correspondingly larger. 

Fig. \ref{dqdt} reveals some rather unexpected compositional trends with time, despite the large scatter of retrieved parameters in Fig. \ref{comp}.  Poleward of 60$^\circ$S, both acetylene and ethane show a decreasing abundance with time, broadly consistent with the trends identified over much of the southern hemisphere by \citet{13sinclair} using 15-cm$^{-1}$ resolution CIRS spectra.  However, the depletion is far stronger within 5$^\circ$ of the south pole above the south polar cyclone, suggesting that the peak in hydrocarbon abundances observed at the south pole in Fig. \ref{comp} is being depleted faster than the abundance at any other latitude, by $0.030\pm0.005$ ppm/year and $0.35\pm0.1$ ppm/year for C$_2$H$_2$ and C$_2$H$_6$ at 1 mbar, respectively.  

These south polar hydrocarbon depletions are mirrored by a general enhancement in hydrocarbon abundances at the north pole in Fig. \ref{dqdt}, although the changes are not localised to within 5$^\circ$ of the pole, but are instead distributed over a wider area.  Ethane has increased by $0.45\pm0.1$ ppm/year at 1 mbar poleward of 77$^\circ$N, whereas lower latitudes show an increase of only $0.10\pm0.03$ ppm/year at 1 mbar.  The boundary between these two regions is rather sharp, potentially associated with the prograde tropospheric jet at this latitude (also the location of Saturn's northern hexagon).   The behaviour of acetylene is rather different, with $\pderiv{q}{t}$ varying smoothly from low values near 70-75$^\circ$N ($0.005\pm0.005$ ppm/yr) to higher values near 85$^\circ$N ($0.030\pm0.003$ ppm/yr), with no evidence of a sharp gradient associated with the prograde jet.  These changes in the $\pderiv{q}{t}$ values are consistent with the trends observed in Fig. \ref{compchange}, and the different behaviours of the two hydrocarbons, which may trace the general stratospheric circulation, will be discussed in Section \ref{discuss_cxhy}.

\section{Discussion}
\label{discuss}

\subsection{Comparison with radiative climate models}
\label{compare}
The polar atmospheric temperature changes shown in Section \ref{Tres} are governed by the spatial and temporal distributions of radiative heaters and coolants, combined with vertical and horizontal circulations that generate adiabatic heating and cooling that act to mitigate temperature differences.  With no direct method for assessing the strength of the general circulation, we utilise radiative climate models to isolate the components driven by radiative effects.    Seasonal radiative models balance (i) the radiative heating due to sunlight absorption by methane and aerosols with (ii) radiative cooling via mid-infrared emission from ethane and acetylene (and to a lesser extent, methane) in the stratosphere, and collision-induced continuum emission from H$_2$-H$_2$, H$_2$-He and H$_2$-CH$_4$ in the upper troposphere.  Radiative equilibrium models have increased in the sophistication of their handling of radiative transfer and opacity sources since those early models that were developed to explain the thermal asymmetries observed by Voyager/IRIS \citep{85bezard, 90conrath, 92barnet}.  The spatial distributions of aerosols and hydrocarbons are typically assumed to be latitudinally-uniform \citep{12friedson, 14guerlet}, which may lead to offsets in predicted seasonal amplitudes.  \citet{08greathouse_agu} and \citet{10fletcher_seasons} used a latitudinal distribution of hydrocarbons derived from a single epoch \citep[typically southern summer,][]{05greathouse,09guerlet} and fixed these for the entire Saturnian year.  However, these hydrocarbon distributions lack spatial resolution in the polar regions, being relatively flat poleward of 60$^\circ$, counter to the findings in Section \ref{compres}.  Despite these differences in the handling of opacity sources, these models are capable of reproducing the broad trends observed in CIRS data \citep{10fletcher_seasons, 14guerlet} in the absence of dynamics.  Eddy mixing and advective transport are combined with radiative equilibrium calculations in the model of \citet{12friedson}, but they caution that their simulations of high latitudes may not be robust.

\begin{figure*}[htbp]
\begin{center}
\includegraphics[width=18cm]{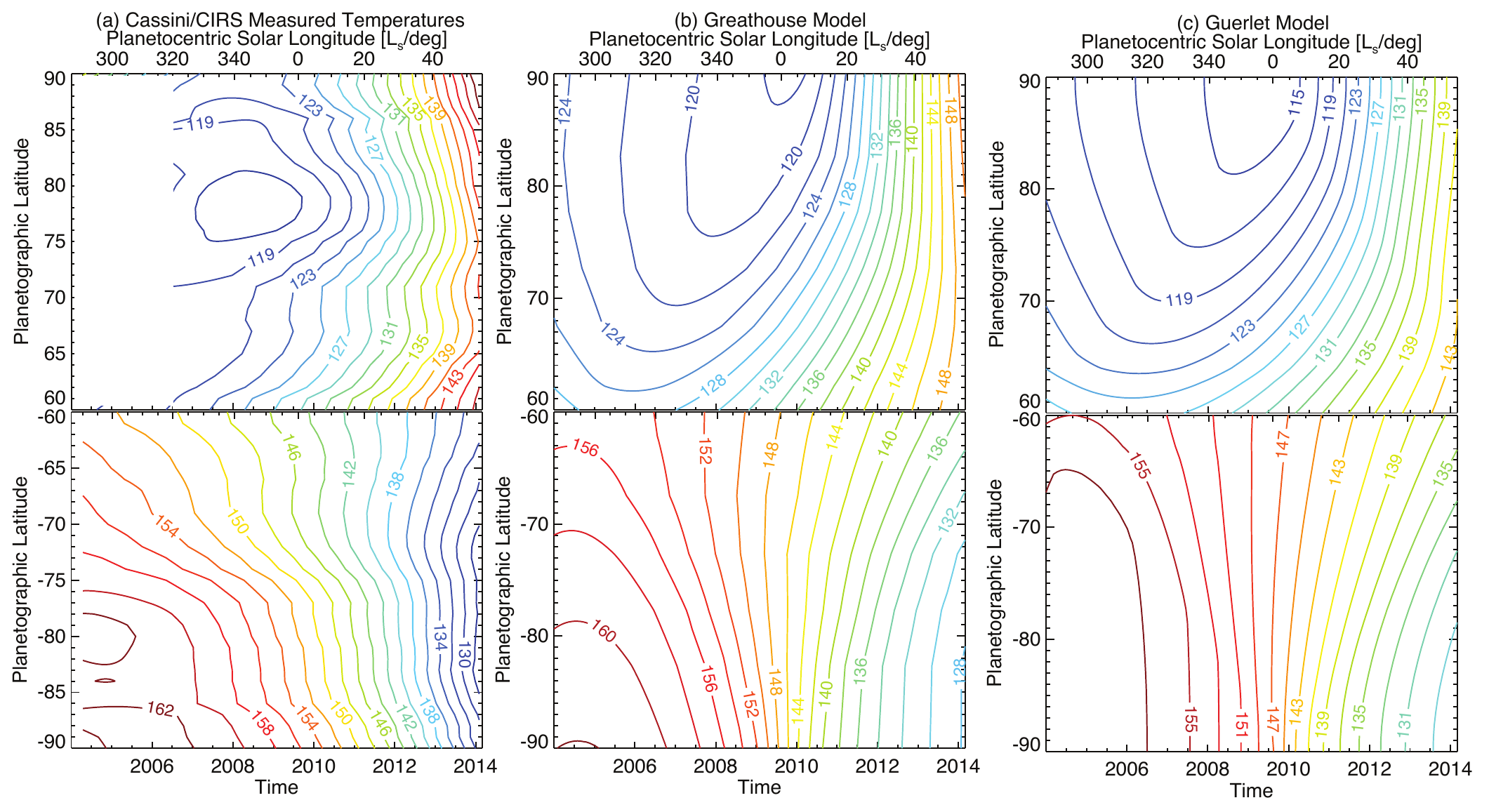}
\caption{Comparing the CIRS-derived 1-mbar temperature fields (column a) for the north and south pole with the predicted temperatures from the radiative climate models of \citet{08greathouse_agu} and \citet{14guerlet}.  Although localised differences are apparent due to stratospheric circulation, the models are able to reproduce the trends observed in the dataset.  See the caption for Fig. \ref{Trecons} for a discussion of uncertainties on the measured temperatures.}
\label{model_1mb}
\end{center}
\end{figure*}

In Fig. \ref{model_1mb} we compare the 1-mbar temperature fields at the north and south poles derived from CIRS results (e.g., using the quadratic fits to the measured trends) with the predictions of two recent radiative climate models.  Unlike the model of \citet{14guerlet}, which features an internal heat source, convective readjustment and tropospheric aerosols to simulate tropospheric temperatures, the model of \citet{08greathouse_agu} lacks tropospheric aerosols or the opacity of PH$_3$, so is not applicable for altitudes below the 10-mbar level.  In this instance, the Greathouse model utilises the hydrocarbon distributions derived by \citet{09guerlet}, rather than the elevated abundances derived by \citet{05greathouse}, which was shown to produce a 3-5 K offset in the predicted 1-mbar temperatures \citep{10fletcher_seasons}.  The effects of ring shadowing are included in both models but are not relevant at such high latitudes.  Comparing Fig. \ref{model_1mb}(a) with panels (b) and (c) shows that the models reproduce the gross structure - the timing of the minimum north polar temperatures gets closer to spring equinox for increasing northern latitude, and the southern hemisphere shows a cooling at all latitudes.  The models do not reproduce (i) the increase in stratospheric temperature poleward of 75-80$^\circ$N; and (ii) the gradients in stratospheric temperature observed near 75$^\circ$S and 87$^\circ$S, which are dynamically associated with zonal jets in the troposphere.  

Given that the hydrocarbons are fixed with time and are not elevated at high latitudes in these models, both reproduce the range of observed temperatures remarkably well.   At the north pole, temperatures range from 116-150 K, which is close to the 118-150 K range predicted by the Greathouse model and similar to the 114-143 K range predicted by the Guerlet model.  Inspection of the results of \citet{12friedson} shows a 120-150 K range at 80$^\circ$N over this time period, consistent with the CIRS results.  At the south pole the 128-164 K range measured by CIRS is reproduced in the Greathouse model (128-162 K), but the peak temperatures are slightly underestimated by the Guerlet model (128-158 K).  \citet{12friedson} suggest a range at 80$^\circ$S of 140-160 K over this period, with less cooling than that observed.  In general, the Guerlet model predicts 1-mbar temperatures slightly cooler than those measured, although we have seen that different assumptions about the hydrocarbon distributions can lead to small offsets in the predicted temperatures.  This would be particularly true if high-altitude hydrocarbon-enriched air (with shorter radiative time constants) was advected downwards.  Finally, these models lack stratospheric aerosols, which could provide a significant component in the radiative budget at the high latitudes.

\begin{figure*}[htbp]
\begin{center}
\includegraphics[width=13cm]{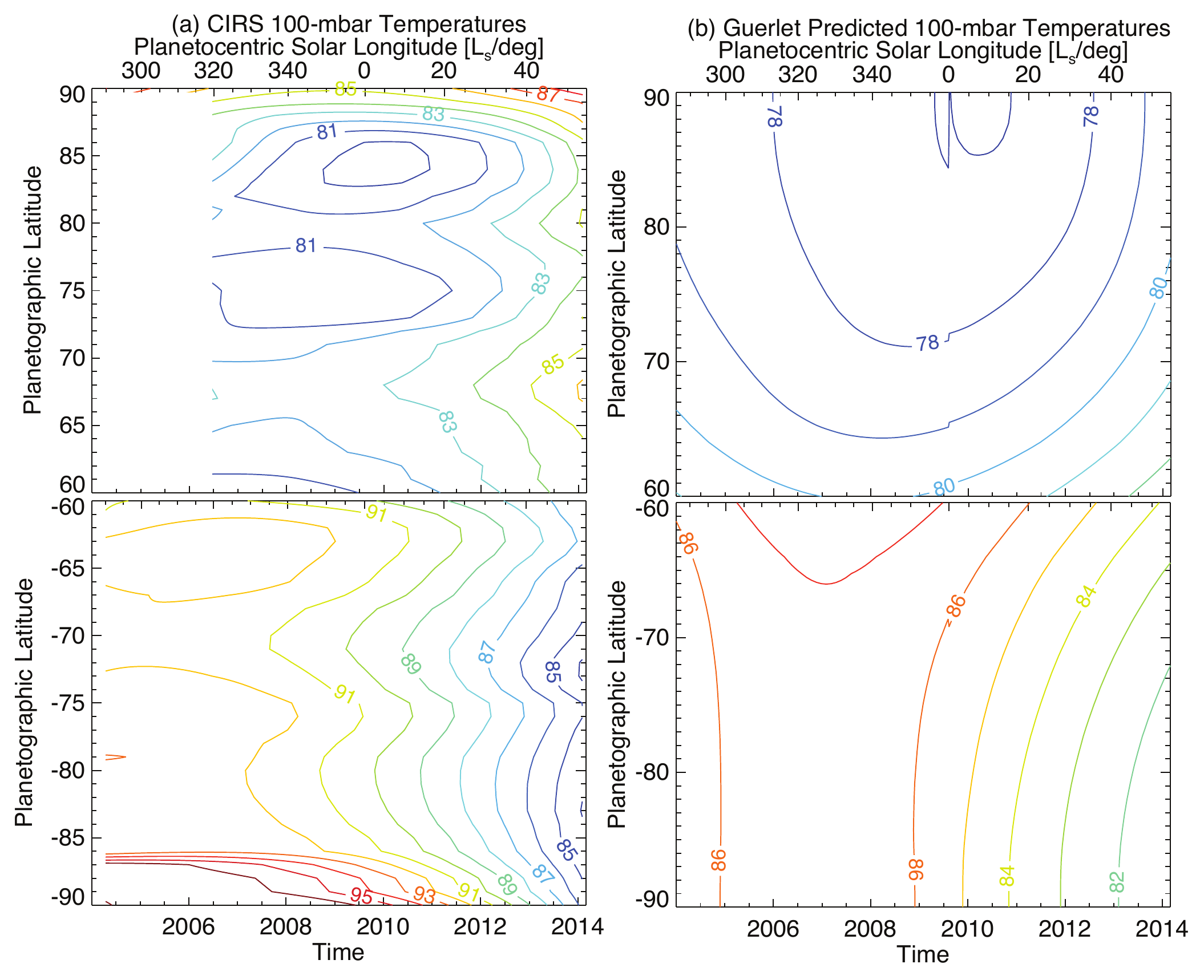}
\caption{Comparison of the measured 100-mbar temperature changes with those predicted by the radiative climate model of \citet{14guerlet}.  The primary differences arise from dynamic perturbations of the temperature field, but the gross structures are similar (e.g., timing of temperature minima and maxima).  See the caption for Fig. \ref{Trecons} for a discussion of uncertainties on the measured temperatures.}
\label{model_100mb}
\end{center}
\end{figure*}

Figs. \ref{dTdt}(b-c) are a direct comparison of the model to the measured $\pderiv{T}{t}$ in Fig. \ref{dTdt}(a), simply using a linear fit to the modelled $\pderiv{T}{t}$ at each latitude and altitude.  Clearly a linear assumption is invalid in cases where a temperature minimum or maximum is present within the 10-year timespan of observations, and the comparison in Fig. \ref{dTdt} should be treated solely as a means of visualisation.  Fig. \ref{dTdt} shows that the approximate rate of the temperature change (5 K/yr) is reproduced by both models, albeit without the distinct regions of excess heating or cooling poleward of 75$^\circ$ in Fig. \ref{dTdt}(a).  The Greathouse model predicts no thermal evolution for $p>10$ mbar, whereas the Guerlet model (with the additional tropospheric opacities) predicts small seasonal changes ($<1$ K/yr) in the upper troposphere, which is more consistent with the CIRS observations.  The 100-mbar temperatures predicted by the Guerlet model are compared to the CIRS measurements in Fig. \ref{model_100mb} - the key difference arises from the dynamically-forced belt/zone structure and the localised polar cyclones in the data, but in general the retrieved temperatures are warmer than those predicted by the model by 1-2 K in the northern hemisphere and 5-6 K in the southern hemisphere.   The timing of the maximum and minimum temperatures, and the magnitude of $\pderiv{T}{t}$ are reasonably consistent between model and data.

We have seen that radiative climate models produce seasonally-varying temperature fields that are qualitatively consistent with the CIRS observations, indicating that the radiative balance between methane and aerosol heating and hydrocarbon cooling is relatively well understood, despite the time-invariant composition in these models.  We expect that the use of time-dependent hydrocarbon distributions, such as those provided in Section \ref{compres}, will help to refine these values.  Nevertheless, there remain significant departures from the radiative models, and we must invoke either vertical motions (adiabatic heating and cooling, see Section \ref{discuss_vert}) or additional heaters and coolants (e.g., aerosols) that are not currently accounted for in these models.

\subsection{Implications for atmospheric circulation}

\subsubsection{Vertical Motions}
\label{discuss_vert}

Atmospheric subsidence is often invoked as an explanation for the warm polar cyclones (which have been stable throughout the Cassini mission at both poles, Fig. \ref{polarmap}) and the larger, warm stratospheric hood in the summer hemisphere \citep{05orton, 05flasar, 08fletcher_poles, 09dyudina}.  By analogy to Jupiter's 5-$\mu$m hotspots, where cloud- and volatile-free conditions are produced by localised downdrafts in the tropospheric cloud decks, the polar vortices are thought to be regions of subsidence and adiabatic heating, isolated from the rest of the atmosphere by peripheral zonal jet streams.  \citet{08fletcher_poles} confirmed that the south polar cyclone was bounded by the jet at 87$^\circ$S and suggested that a similar prograde jet would be discovered at 88$^\circ$N, bounding the north polar cyclone.  This was later confirmed by \citet{09baines_pole} using nighttime cloud tracking at 5 $\mu$m from Cassini/VIMS in 2008, who showed that the tropospheric winds drop with latitude poleward of the cyclone wind maximum of $136\pm6.5$ m/s.  The `eyes' of the cyclones are cloud-free, surrounded by concentric eyewalls and rings of high-altitude hazes \citep{08dyudina, 09dyudina}.   Thus, in the troposphere at least, the warm cyclone temperatures are deeply connected with the zonal flow (and, by mass conservation, with the upper tropospheric overturning). 

\begin{figure*}[htbp]
\begin{center}
\includegraphics[width=14cm]{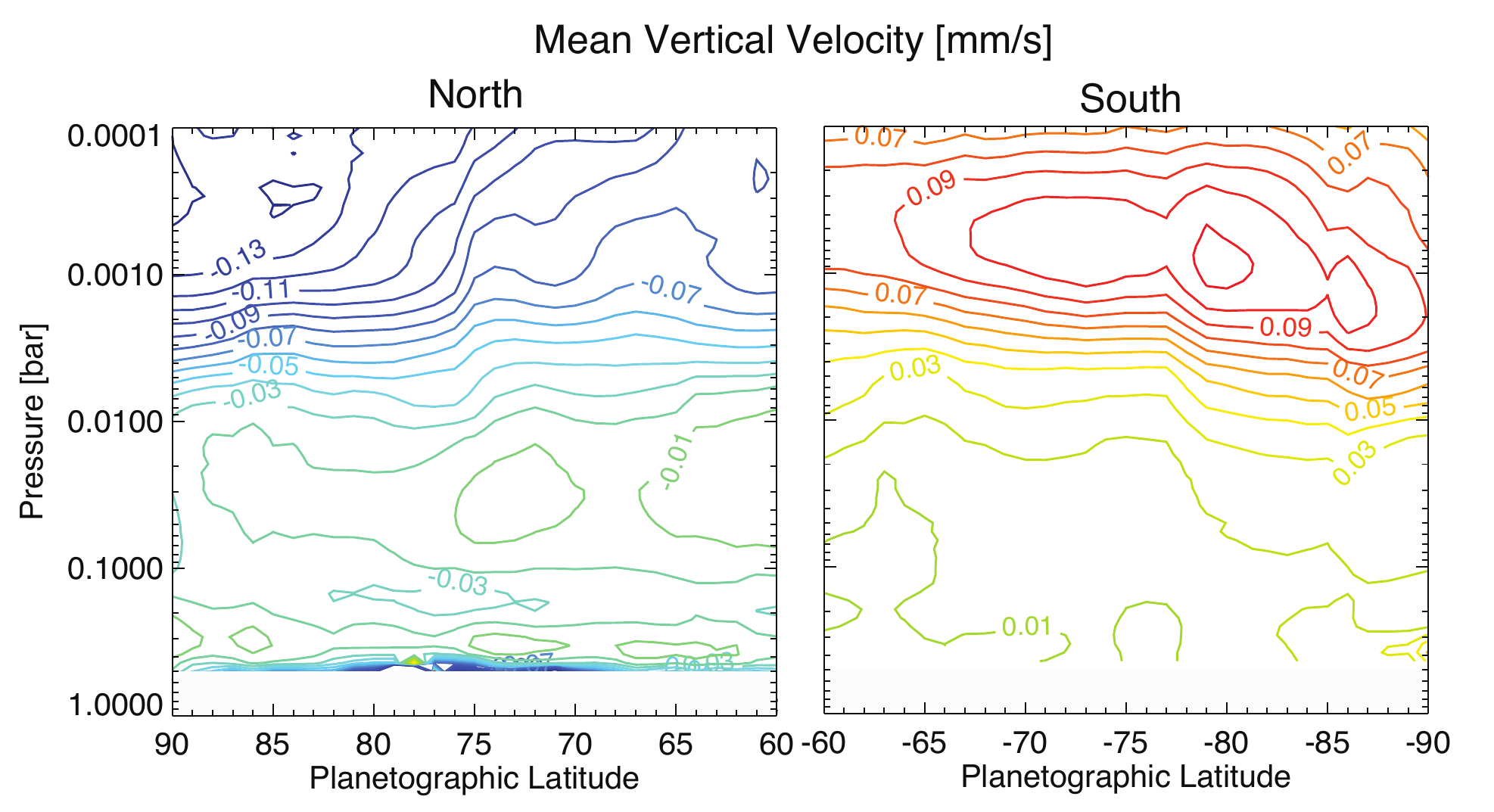}
\caption{Residual mean vertical circulation in mm/s estimated via the thermodynamic energy relation, containing contributions from the temperature anomaly \citep[the difference between the retrieved $T(p)$ and the radiatively-forced temperatures of][]{14guerlet} and the measured rate of change of temperature with time (Fig. \ref{Trecons}).  Positive values of $w$ imply upwelling, negative values imply subsidence.  Random uncertainties in $w$ are approximately 17\% near 1 mbar and 25\% near 100 mbar, although this does not account for systematic offsets.}
\label{vert_vel}
\end{center}
\end{figure*}  

We can use the retrieved temperature variability to estimate the sense and direction of the atmospheric circulation, assuming that the radiative budget of the atmosphere is well understood.  In the absence of phase changes and any frictional energy dissipation, the change in the temperature field with time (given by the advective derivative of temperature, $DT/Dt$) is related to vertical motions causing adiabatic heating and cooling and the net radiative heating ($Q$).  This leads to the thermodynamic energy relationship \citep{03hanel, 04holton} in terms of the mean residual meridional circulation:
\begin{equation}
\left(\pderiv{T}{t}+w\left[\pderiv{T}{z}+\frac{RT}{Hc_p}\right]\right) = \frac{Q}{\rho c_p} \approx \frac{T_E-T}{\tau}
\label{eq_heat}
\end{equation}
where we assume that zonal and meridional motions have no contribution to the advective derivative of temperatures, $DT/Dt$.  Zonal advection of heat is omitted as the temperature is assumed to be longitudinally homogeneous, and meridional advection is ignored as the vertical potential temperature gradient is so much larger than the horizontal.  $w$ is a zonally-averaged approximation to the residual mean vertical windspeed, $H$ is the atmospheric scale height, $R$ is the molar gas constant divided by the mean molar weight of Saturn's atmosphere and $c_p$ is the specific heat capacity of the atmosphere with density $\rho$.  The specific heat capacity is calculated for every pressure level assuming a simple H$_2$-He-CH$_4$ atmosphere and assuming a `frozen' 3:1 ratio for the ortho-to-para-H$_2$ fraction. Note that the heating effects due to the lagged conversion between ortho- and para-H$_2$ are not accounted for \citep{98conrath}. 


The final approximation on the right hand side of Eq. \ref{eq_heat} relates the radiative heating rate to the temperature anomaly: the seasonally varying (but diurnally averaged) instantaneous radiative equilibrium temperature ($T_E$) \citep{90conrath}, minus the retrieved temperature ($T$) and divided by the radiative time constant ($\tau$) provided by \citet{90conrath}.  Using the hydrostatic equation to convert to log-pressure coordinates, we find that the vertical velocity can be expressed as:
\begin{equation}
w \approx H \left[  \frac{\frac{T_E-T}{\tau} - \pderiv{T}{t}}{\frac{RT}{c_p} - \pderiv{T}{\ln{p}}} \right]
\end{equation}
The residual mean velocity is therefore comprised of two terms - an `anomaly term' related to the deviation of the temperature from the instantaneous radiative equilibrium value and a `temporal term' related to the rate of change of temperature with time.  We approximate the instantaneous radiative equilibrium by the radiatively forced temperatures interpolated from the model of \citet{14guerlet}, which should be a reasonable approximation in the stratosphere where infrared cooling to space dominates, but not necessarily in the troposphere where the model includes convective re-adjustment and an internal heat flux.  

Calculating $w$ for every latitude, pressure and date during the Cassini mission, and then averaging over the ten years of observations, we show the zonally- and temporally-averaged mean vertical velocities in Fig. \ref{vert_vel} (the temporal average hides the fact that $w$ is actually varying with time as the poles undergo warming and cooling).  The `anomaly' component of $w$ is generally smaller than the `temporal' component, such that the warm polar temperatures (suggestive of $w<0$ mm/s at both poles) are actually a secondary effect compared to the northern warming trend ($w\approx -0.1$ mm/s) and southern cooling trend ($w\approx+0.1$ mm/s).  Although systematic uncertainties are difficult to quantify (see below), we can estimate random uncertainties by propagation of (i) errors on $\pderiv{T}{t}$ from the smooth temporal interpolation (e.g., Fig. \ref{Trecons}), providing approximately 0.003-0.005 mm/s (or 12-15\%) to the uncertainty in the `temporal' component of $w$; and (ii) errors on $T_E-T$ from the retrieved $T(p)$ uncertainty, providing 0.006 mm/s uncertainty to the `anomaly' component near 1 mbar (5-8\%) and 0.003 mm/s near 100 mbar (10-20\%). Adding in quadrature, this provides random uncertainties on $w$ of 17\% at 1 mbar and 25\% at 100 mbar.  The latitudinal variations observed in Fig. \ref{vert_vel} are therefore significant in the stratosphere, but not in the troposphere.  

The magnitude of $w$ can be compared to the estimate of \citet{05flasar} from early in the mission ($w\approx -0.1$ mm/s at the south pole in 2004), who neglected the `anomaly' component (i.e., they assumed that the radiative timescales are sufficiently long that the net diabatic heating could be considered negligible) and assumed that south polar temperatures were continuing to warm.  Hence, given our more precise knowledge of $\pderiv{T}{t}$ and an estimate of the anomaly from radiative equilibrium, we disagree with the direction of motion, but agree with their magnitude.  Furthermore, the magnitude of our estimated winds compares favourably with the 1-mbar model output of \citet{12friedson}, albeit at lower latitudes. 

Fig. \ref{vert_vel} shows that the vertical velocities vary with latitude and altitude, being largest in the 0.1-5.0 mbar range at both poles ($w\approx-0.15$ mm/s in the northern stratosphere, and $w\approx+0.11$ mm/s in the southern stratosphere).  The vertical location is possibly due to the absence of temperature information for $p<0.5$ mbar in the CIRS nadir data, and vertical motions are expected to persist and possibly strengthen at higher altitudes \citep{12friedson}.  Indeed, vertical velocities outside of the 0.5-5.0 mbar and 70-250 mbar sensitivity ranges of the CIRS data are estimated from the interpolated temperature fields and should be viewed with caution.  At the north pole, the subsidence velocities show a marked increase poleward of 75$^\circ$N, mirrored by an increase in southern hemisphere upwelling poleward of 77$^\circ$S.  Velocities tend towards zero in the deeper atmosphere where temperature variability is negligible, except within the polar cyclonic vortices themselves - at 100 mbar we predict velocities of $w\approx-0.02$ mm/s in the NPC and $w\approx+0.02$ mm/s in the SPC (comprised of an upwelling component from the `temporal' term and a subsiding component from the `anomaly' term), of the same order of magnitude as that proposed by \citet{08fletcher_poles} to explain the polar hotspots.

The precise value of $w$ depends upon many assumptions:  the omission of meridional advection from the heat equation (Eq. \ref{eq_heat}); the `frozen equilibrium' assumption in calculating the heat capacity; the validity of using temperature deviations from the model of \citet{14guerlet} as an approximation to the net diabatic heating.  Systematic offsets between their radiative convective model and our retrieved $T(p)$ (e.g., Fig. \ref{model_1mb}) could add a systematic offset to the `anomaly' term, and hence the mean $w$.  For example, the existence of latitudinally-variable aerosols within the polar regions may contribute substantially to the radiative budget, increasing the heating and cooling rates compared to the case without aerosols, therefore altering the size of the offset from equilibrium.  The likely contribution of these polar hazes (potentially from condensed hydrocarbons) to the radiative budget is supported by the observation that Saturn's aerosols change character within the polar regions.  Stratospheric particles appear to be smaller \citep[0.1 $\mu$m radius at high latitudes compared to 0.2 $\mu$m at low latitudes,][]{05perez-hoyos}, more UV-absorbent \citep{93karkoschka} and with an increased optical depth poleward of $\pm70^\circ$ \citep{01stam, 05karkoschka}.  Radiative transfer modelling by \citet{06sanchez} suggests that the haze within the south polar stratosphere resides between 1-30 mbar, precisely the region probed by the CIRS data.  In the absence of stratospheric aerosols from the radiative climate models (particularly at the highest latitudes), the estimates of $w$ must be considered as approximations at best, and in Section \ref{cxhy_temporal} we seek to confirm them using the changing hydrocarbon distributions.   




\subsubsection{Horizontal Motions}
\label{winds}

\begin{figure*}[htbp]
\begin{center}
\includegraphics[width=18cm]{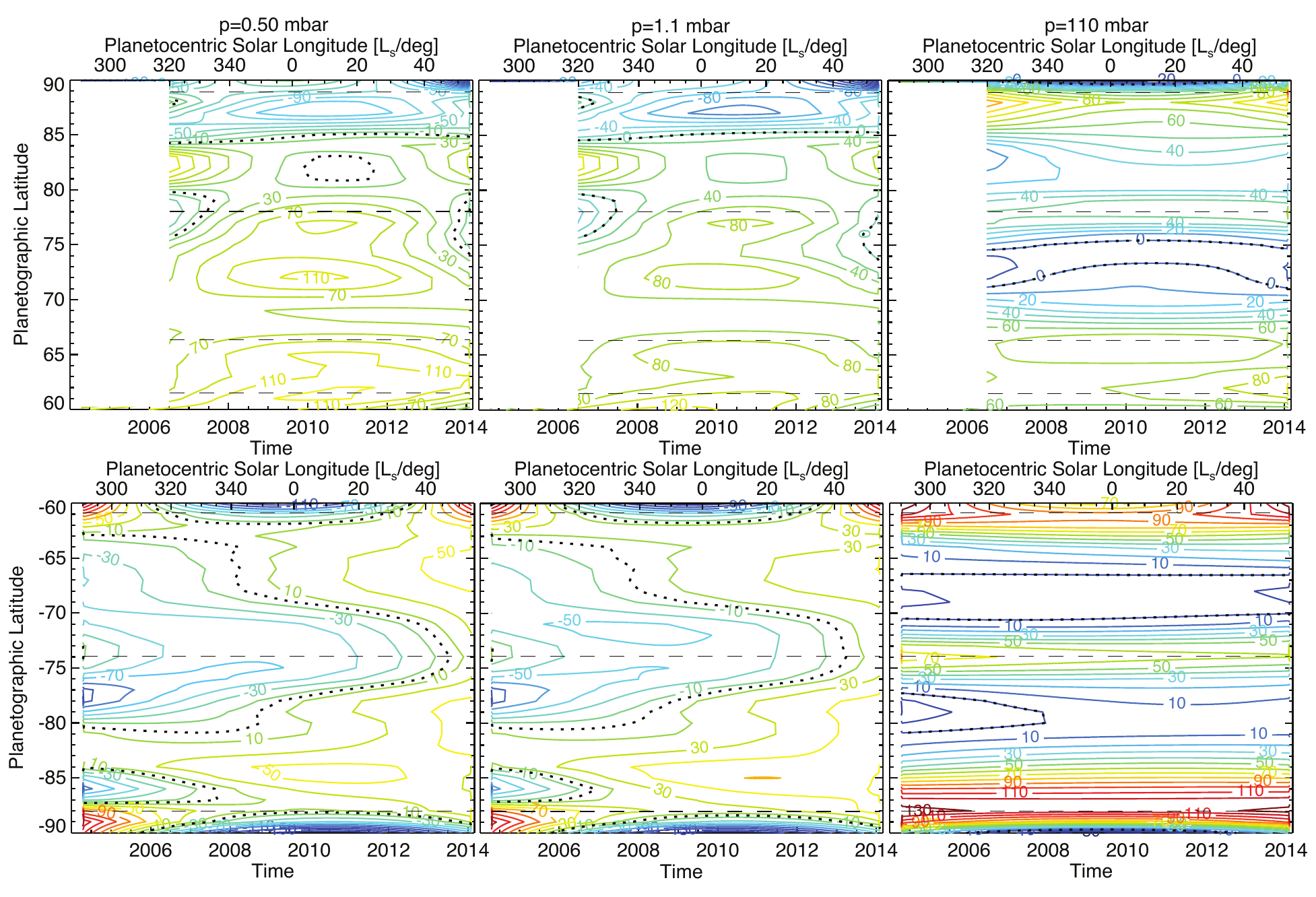}
\caption{Zonal winds estimated for three atmospheric levels in the northern (upper panels) and southern (lower panels) polar regions.  Windspeeds in m/s were estimated by integrating the thermal windshear relationship for all dates and latitudes to show how the zonal jets are expected to respond to the seasonal changes.  The absolute values assume no uncertainty in the latitudinal derivative of temperature.  Horizontal dashed lines show the peak prograde jet locations in the troposphere for reference.  The black dotted contour is the zero wind level, indicating locations where the zonal wind direction flips from prograde ($u>0$ m/s) to retrograde ($u<0$ m/s).  Northern hemisphere wind estimates prior to mid-2006 are omitted (e.g., Fig. \ref{coverage}).  }
\label{zonalwind}
\end{center}
\end{figure*}

In addition to vertical motions, the shifting meridional temperature gradients are related to the vertical shear of the zonal winds via the thermal windshear relation, assuming geostrophic balance between the meridional pressure gradients and Coriolis forces.  In log-pressure coordinates, the thermal windshear equation for the zonal ($u$) direction is:
\begin{equation}
f\pderiv{u}{\ln \left( p \right)} = \frac{R}{a} \pderiv{T}{\psi} = R\pderiv{T}{y}
\end{equation}
where $T$ is the temperature in Kelvin, $p$ is the pressure in bar, $f$ is the Coriolis parameter $f=2\Omega \sin(\psi)$ (where $\Omega$ is the planetary angular velocity, $\psi$ is the planetographic latitude), and $a$ is the planetary radius at the relevant latitude, accounting for Saturn's oblateness.  We use the temperature fields reconstructed with smooth polynomials from the noisy ensemble of retrievals, and calculate the changing $\pderiv{u}{\ln(p)}$ with time.  Zonal wind fields were obtained from continuum-band Cassini imaging \citep{11garcia}, extended northwards with more recent Cassini data (A. Antu\~{n}ano, \textit{personal communication}), and the peaks of the prograde tropospheric zonal jets occur at planetographic latitudes of 88.9, 78.0 and 66.3 and 61.5$^\circ$N in the northern hemisphere and 60.9, 73.9 and 88.0$^\circ$S in the southern hemisphere \citep{11garcia, 09baines_pole}.  Note that the northern polar region seems to feature four prograde jets, whereas the southern hemisphere has only three, which is consistent with the measured thermal gradients. These zonal winds were assumed to reside at the 500-mbar level, and used to integrate the thermal windshear relation to provide the zonal winds in Fig. \ref{zonalwind}.  Absolute wind values at each altitude are extremely sensitive to the uncertainty in $\pderiv{T}{y}$ - indeed, uncertainties estimated from the standard deviation of the derivative completely mask any of the temporal variability observed in Fig. \ref{zonalwind}, given the 1.0-3.0 K $T(p)$ uncertainties and the error growth as we integrate with height. Nevertheless, $\pderiv{T}{y}$ is certainly changing, so $\pderiv{u}{\ln(p)}$ must also be varying with season.  By treating the reconstructed temperature fields in an identical fashion for every latitude and date (i.e., assuming no uncertainty in $\pderiv{T}{y}$), Fig. \ref{zonalwind} can still be used to discuss general trends in zonal motions even if absolute values remain highly uncertain. 

Fig. \ref{zonalwind} suggests that the upper tropospheric winds are relatively stable, consistent with the small seasonal gradients identified at $p>0.1$ bar.  The locations of maximal windshear match the jet locations precisely in the southern hemisphere, but the picture in the northern hemisphere is more confused, with the peak 100-mbar jet existing 1$^\circ$ further south in the upper troposphere than it does in the cloud tops \citep{09baines_pole}.  Zonal jet variations of order 10 m/s are suggested in the broad southern retrograde regions, with the jet at 79$^\circ$S varying from weakly retrograde at the start of the mission to weakly prograde by 2014, whereas the jet at 68$^\circ$S remains weakly retrograde for the 10-year duration.  This is consistent with the shifting gradients observed near 100 mbar in Fig. \ref{Trecons}.  The northern tropospheric jets are more variable, with both the 78$^\circ$N prograde jet and the 73$^\circ$N retrograde jet appearing to move northwards between 2004 and 2011, before a slow southward motion to 2014.  However, the absence of cloud tracers near 110 mbar prevents corroboration of these jet motions, which are determined only indirectly from the thermal field.  

Confirmation of the stratospheric wind velocities and variability is even harder, without any clouds to trace the zonal jets.  Nevertheless, the stratospheric $\pderiv{T}{y}$ has changed substantially with season, tending to promote positive vertical shear in the winter hemisphere and negative vertical shear in the summer hemisphere \citep{12friedson}.  The 1-mbar retrograde jets near 74$^\circ$S and 78$^\circ$N (associated with prograde jets in the troposphere and defining the edges of the wide stratospheric vortices) have both weakened and then become prograde as the seasons progressed (the transition from retrograde to prograde occurred in 2012-2013 in the southern hemisphere and 2007-2008 in the northern hemisphere at 1 mbar).  In both hemispheres this corresponds to an increasingly positive $\pderiv{T}{y}$, but in one case this is a weakening of the south polar temperature contrast, and in the other a strengthening of the north polar temperature contrast.   The picture is different at the boundaries of the smaller polar cyclones - in the south, $\pderiv{T}{y}<0$ so that the tropospheric prograde jet remains prograde in the stratosphere throughout the dataset; in the north $\pderiv{T}{y}>0$ and strengthens with time so that the prograde stratospheric jet weakens and becomes more retrograde with time at 1 mbar.    The correspondence between stratospheric temperature gradients exerting shear on the tropospheric zonal jets is not clear cut, and these velocities remain to be tested by direct measurements (e.g., Doppler shifting of stratospheric emission lines).

In summary, as we expect the zonal wind field to remain in balance with the large-scale temperature contrasts (and the resulting vertical shears) as they change, the middle-atmospheric wind field may also be seasonally variable.  The zonal wind profile in the upper troposphere and lower stratosphere is important to understand in the context of vertically propagating waves, as the ability of the atmosphere to trap these waves vertically will be changing over time and influencing the dynamics of Saturn's middle atmosphere.

\subsection{Hydrocarbons}
\label{discuss_cxhy}
\subsubsection{Spatial Variations}

The latitudinal variations of the hydrocarbons in Fig. \ref{comp} have been identified in previous studies of CIRS and ground-based data.  The equator-to-pole decrease in acetylene (C$_2$H$_2$), which can be seen from 60-80$^\circ$S in Fig. \ref{comp} and tentatively from 60-80$^\circ$N, was observed at lower spatial resolution by \citet{05greathouse, 07howett, 09guerlet, 09hesman}.  The same is true of the relatively uniform ethane (C$_2$H$_6$) distribution over the southern hemisphere.   These studies, biased towards southern summer and the early stages of the Cassini mission, lacked significant coverage of the cold northern hemisphere.  \citet{13sinclair} extended the latitudinal studies into the northern hemisphere with a higher spatial resolution dataset, and identified a clear acetylene minimum at 80$^\circ$S followed by a sharp rise towards the south pole, as shown in Fig. \ref{comp}a.  Similar, shallower latitudinal gradients are also observed in the distribution of ethane towards the south pole in Fig. \ref{comp}b.  These polar enhancements within $\pm5^\circ$ of the pole could have been caused by the subsiding branch of Saturn's stratospheric circulation (i.e., the `anomaly term' in Section \ref{discuss_vert}), transporting hydrocarbon-enriched air down from the source regions at microbar pressures.  The peak abundances, coinciding with the warmest temperatures, occur right at the south pole.

In the northern hemisphere, \citet{13sinclair} observed a rise in acetylene and a fall in ethane towards the north pole, albeit with low confidence because of the cold stratospheric temperatures during their 2005-2010 observations.  Fig. \ref{comp}a supports a steep acetylene rise towards the north pole, but only in post-equinox observations (i.e., when the stratosphere was warming).  Before equinox, the acetylene distribution appears much more uniform.  The spatial distribution of ethane in Fig. \ref{comp}b suggests a shallow rise towards the north pole in post-equinox observations, and a relatively uniform polar abundance before equinox, in disagreement with the polar decrease observed by \citet{13sinclair}.  Spatial gradients in hydrocarbon species are hard to identify in north polar data because of the low signal (i.e., cold stratospheric temperatures) present there during most of the dataset, but should become clearer as the temperatures rise towards northern summer solstice.   

\subsubsection{Temporal Variations}
The seasonal variability of the two principal products of methane photolysis (ethane and acetylene) is expected to have a substantial effect on the radiative budget in Saturn's stratosphere, given that these species are the dominant coolants \citep{85bezard}.  \citet{05moses_sat} developed a time-dependent 1D diffusive photochemistry model at multiple latitudes, accounting for vertical mixing (via a mixing coefficient $K_{zz}$), ring shadowing, insolation and solar cycle variability to explore the chemical changes, but neglecting any horizontal mixing (coefficient $K_{yy}$) of these species.  This 1D model assumed a time- and latitude-invariant $T(p)$ structure, using the temperatures determined by ISO in 1996-1997 \citep{01lellouch}, close to Saturn's northern autumnal equinox.  In their model, hydrocarbon production at the south pole peaks near southern summer solstice for microbar pressures, but it takes some time for these changes to diffuse downward.  Hence the peak mole fraction will lag behind summer solstice for deeper and deeper pressures due to the increased vertical diffusion timescales, and for $p>1$ mbar the response times to the changing seasons are expected to exceed a Saturnian year.  Similarly, the minima in hydrocarbon abundances will lag behind the winter solstice.  

\begin{figure*}[htbp]
\begin{center}
\includegraphics[width=16cm]{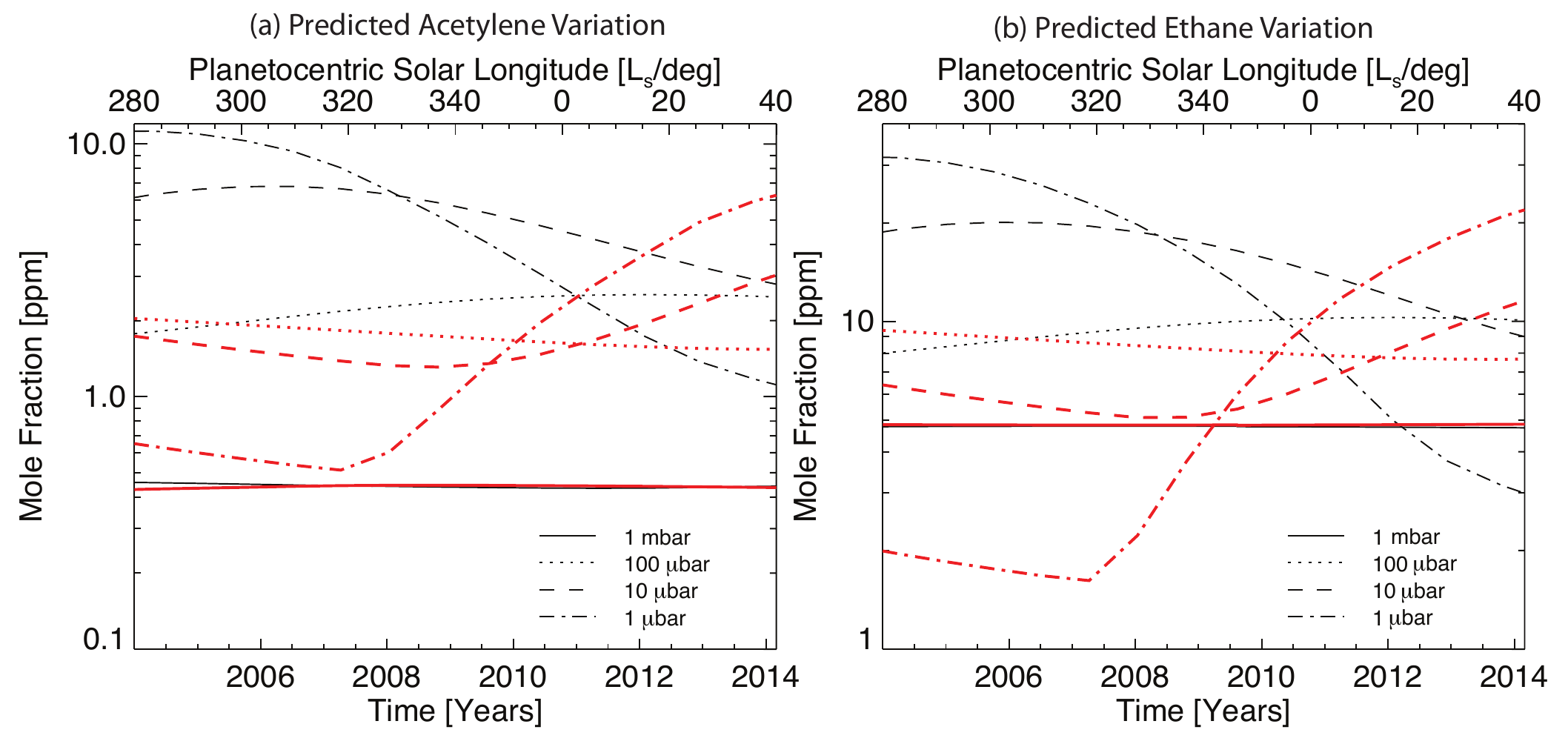}
\caption{Predicted hydrocarbon variability in the updated photochemical model of \citet{07moses}, shown over the timespan of the Cassini observations for 72$^\circ$N (red, thick lines) and 72$^\circ$S (black lines).  Four pressures are shown to demonstrate the magnitude and phase lags of the predicted variability from 1 $\mu$bar to 1 mbar.  No variations are expected at 1 mbar in this simple model.  }
\label{moses}
\end{center}
\end{figure*}

In the present study, we are interested in the observed rate of change of the hydrocarbon distributions, as determined for the polar regions in Fig. \ref{dqdt}.  \citet{07moses} updated the 1D model to permit horizontal coupling between different latitudes, but in this study we use the output of this 2D model with $K_{yy}=0$ (J. Moses, \textit{personal communication}), as this assumes an averaged solar cycle rather than the specific solar cycle variability used by \citet{05moses_sat}.  Fig. \ref{moses} plots the predicted acetylene and ethane variations from the 2D model over the timespan of the Cassini observations for 72$^\circ$N and 72$^\circ$S, representative of the polar regions.  As expected, the model predicts no variability at millibar pressures because the phase lags exceed a Saturnian year, which is inconsistent with the observed $\pderiv{q}{t}$ in Fig. \ref{dqdt}.  Indeed, the variations observed by CIRS appear to track the changes occurring at higher microbar pressures, also shown in Fig. \ref{moses}.  For example, the model predicts that the south polar peak in acetylene and ethane at 0.1 mbar lags the solstice by more than a season, occurring between 2011-2012 at $L_s\approx15^\circ$ (southern autumn).  Similarly, the north polar minimum in hydrocarbons at 0.1 mbar lags winter solstice by more than a season, occurring at $L_s\approx60^\circ$.  Given these predicted variations, we might expect the north polar hydrocarbons near 1 mbar to be decreasing throughout the Cassini dataset, whereas the south polar hydrocarbons should still be increasing.  This is clearly not the case in Figs. \ref{compchange} and \ref{dqdt}, where the measurements reveal trends in the opposite direction.  The south polar maximum and the north polar minimum must have occurred closer to southern summer solstice (i.e., before the start of the Cassini mission).  Indeed, model predictions at 1 $\mu$bar suggest that south polar hydrocarbons should have peaked at $L_s\approx280^\circ$ and north polar hydrocarbons should have been at a minimum near $L_s\approx320^\circ$ in Fig. \ref{moses}, which is broadly consistent with the trends we observe in the CIRS data at much higher pressures (1 mbar).  One possible explanation is that the temporal trends are entirely governed by atmospheric circulation, rather than by the chemical response times to changing insolation, as described in the next section.


\subsubsection{Hydrocarbons as tracers of stratospheric circulation}
\label{cxhy_temporal}

\begin{figure*}[htbp]
\begin{center}
\includegraphics[width=13cm]{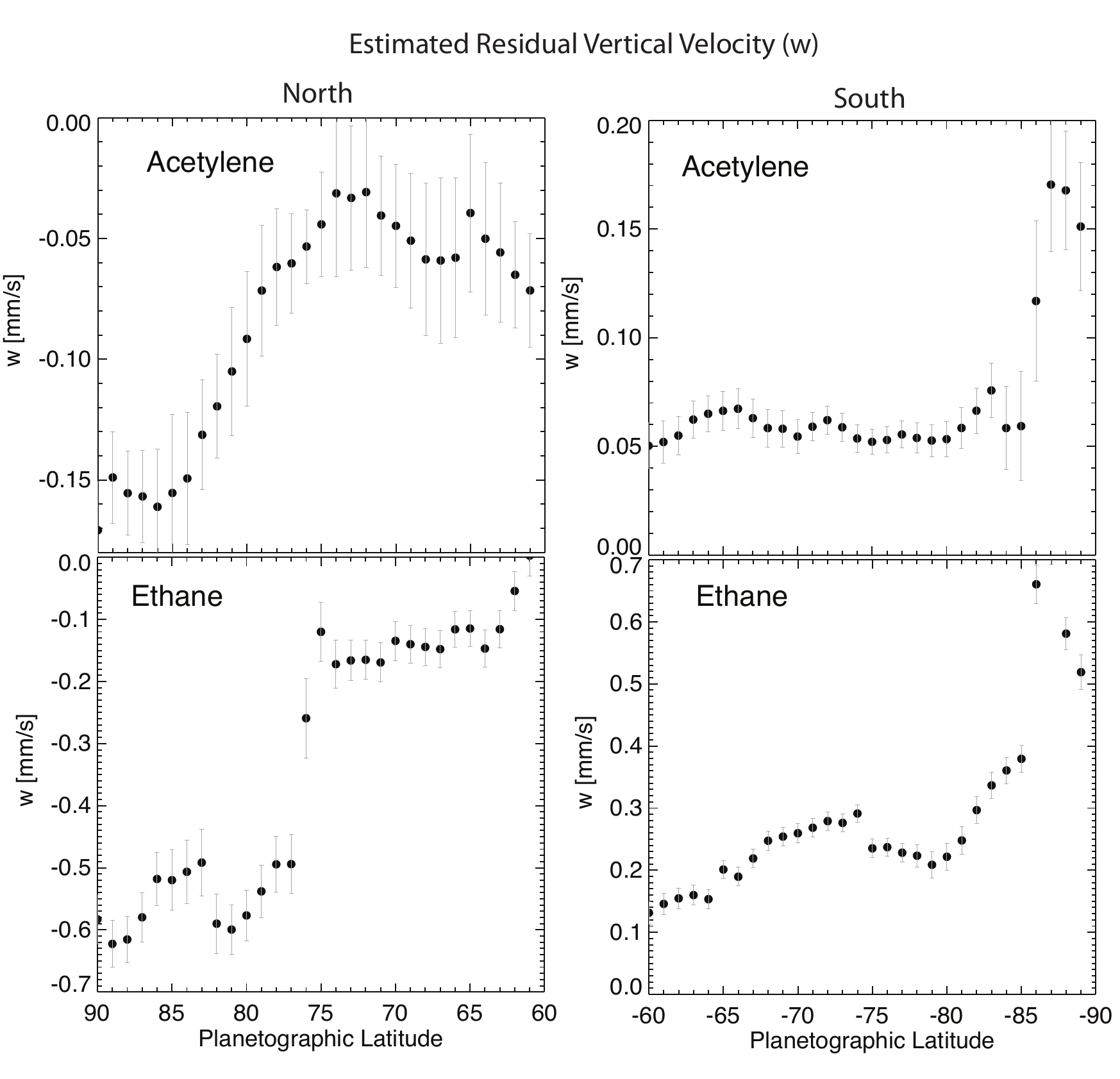}
\caption{Using the temporal evolution of the hydrocarbons in Fig. \ref{dqdt}, we solve the continuity equation for the mean residual circulation $w$ (assuming fixed $\pderiv{q}{z}$ at all latitudes, which is unlikely to be the case).  Uncertainties arise from the errors in $\pderiv{q}{t}$ measured in Fig. \ref{dqdt}. }
\label{comp_vel}
\end{center}
\end{figure*}

We can use the measured $\pderiv{q}{t}$ in a simplistic evaluation of the continuity equation, where the advective derivative of the mole fraction profile ($q(z,t)$) is equated to the sources, sinks ($S$) and eddy transport ($D$) of the chemical species \citep[e.g.,][]{09guerlet};
\begin{equation}
\left(\pderiv{}{t} + w\pderiv{}{z}\right)q = S + D
\end{equation}
where $w$ is the zonally-averaged residual mean vertical circulation.  Meridional advection of hydrocarbons is ignored in this calculation.  Assuming that there are no net sources or sinks of these hydrocarbons (i.e., that production and loss mechanisms have the same efficiency throughout the observational period and thus balance one another precisely, which is reasonable at 1 mbar as shown in Fig. \ref{moses}) and that eddy mixing over the ten-year timespan of the observations is negligible \citep[e.g., the vertical diffusion timescale in Fig. 7 of ][indicate that it is longer than a Saturn year at 1 mbar]{05moses_sat}, we set the right hand side to zero and solve for $w$ (Fig. \ref{comp_vel}, assuming $\pderiv{q}{z}$ from the reference \textit{a priori} profiles).  Fig. \ref{comp_vel} shows that subsidence of $0.15\pm0.02$ mm/s would be required to explain the polar acetylene enhancement (or $0.62\pm0.03$ mm/s for ethane enhancements) over the north pole, whereas upwelling at $0.17\pm0.03$ mm/s for acetylene (or $0.65\pm0.03$ mm/s for ethane) is required to explain the south polar hydrocarbon depletions.  The estimates from acetylene are comparable (in both magnitude and direction) to the vertical motions derived from the thermodynamic energy relation in Section \ref{discuss_vert}, whereas the estimates from ethane are systematically larger.  These crude estimates are consistent with similar vertical mixing rates obtained from global hydrocarbon distributions measured by \citet{09guerlet} and \citet{13sinclair}.  However, our estimates assume that $\pderiv{q}{z}$ is constant with time, which is unlikely to be the case in such a dynamic environment, and may explain why the estimates of $w$ for ethane are inconsistent with those of acetylene.  In reality, we would expect variations in the vertical subsidence rate to influence the hydrocarbon gradients, but measurement of vertical abundance profiles requires limb sounding \citep[e.g., ][]{09guerlet} with higher spectral resolutions than those considered in this study.  

The crude calculations also assume that there are no net sources and sinks of these species other than advective transport, which may not be the case.  For example, the increase in hydrocarbon production that occurred at the north pole at microbar pressures when it emerged into sunlight might have propagated down to millibar pressures faster than expected.  The only way to resolve this degeneracy between chemistry and stratospheric circulation is to pursue the development of coupled chemistry transport models such as that proposed by \citet{07moses}.  Nevertheless, this hydrocarbon study shows a distinct change in the character of stratospheric circulation poleward of 77$^\circ$N associated with the onset of a warm polar vortex in the northern spring, and poleward 85$^\circ$S associated with the continued presence of a warm polar vortex in southern autumn.


%

\section{Conclusions}

The evolving temperatures and stratospheric composition in Saturn's polar regions have been tracked through ten years of Cassini/CIRS 7-16 $\mu$m spectral observations.  All available data poleward of 60$^\circ$ (including spectra where CIRS was not the prime instrument during Cassini's highly inclined orbits, but excluding data with zenith angles exceeding $60^\circ$) were coadded according to latitude, month and spectral resolution (2.5 cm$^{-1}$ and 15.0 cm$^{-1}$).  The spectra were inverted using an optimal estimation retrieval algorithm to derive the vertical temperature structure in the upper troposphere (70-250 mbar) and stratosphere (0.5-5.0 mbar), and scale factors for the vertical distributions of acetylene and ethane.  Smooth quadratic and linear regression curves were used to interpolate over epochs where no Cassini data are available (i.e., when CIRS was viewing from an equatorial vantage point) to reconstruct a continuous record of thermal and chemical variability from northern winter (2004, $L_s=293^\circ$) to northern spring (2014, $L_s=56^\circ$), over a third of a Saturnian year.  This record allows us to draw the following conclusions:

\begin{enumerate}

\item \textbf{Polar cyclones: } The hot polar cyclones (NPC and SPC) discovered within $2-3^\circ$ of both poles during Cassini's prime mission (2004-2008) are persistent features of Saturn's polar tropospheric circulation, being present at all epochs despite changes in the mean temperatures.  The warm air is dynamically isolated from lower latitudes by peripheral zonal jets at 88.9$^\circ$N and 88.0$^\circ$S, respectively, and coincident with the `hurricane-like' clouds observed in reflected sunlight.  The localised warm airmass extends high into the southern stratosphere in summer and autumn, but is not readily apparent in the northern stratosphere in winter and spring. The polar cyclones are surrounded by cool tropospheric polar zones (near $\approx85^\circ$), although the temperature contrast between the SPC and the southern polar zone has increased as summer progressed into autumn.

\item \textbf{Jet boundaries of stratospheric vortices: } Prograde tropospheric jets at 78.0$^\circ$N and 73.9$^\circ$S exist in regions of substantial vertical wind shear, changing from negative shear (and hence retrograde stratospheric jets) in northern winter/southern summer to positive shear (resulting in prograde stratospheric jets) in northern spring/southern autumn.  The transition from retrograde to prograde occurred in 2012-2013 in the southern hemisphere and 2007-2008 in the northern hemisphere at 1 mbar.  These changing vertical shears are expected because the large-scale temperature contrasts and the zonal wind field should remain in balance as the seasons change.  The shifting stratospheric jets may mark the boundaries of the wider stratospheric vortices (i.e., the enhanced emission poleward of 75$^\circ$), isolating them from lower latitudes.  

\item \textbf{Polar belts and hexagon: } Warm tropospheric polar belts exist equatorward of the polar zones (near $\approx85^\circ$), and poleward of the prograde jets at 78.0$^\circ$N and 73.9$^\circ$S.  The visibility of the south polar belt has increased over the duration of the mission as the temperature gradients have changed.  The northern polar belt exhibits the hexagonal structure first observed by \citet{08fletcher_poles}.  The northern hexagon has also persisted throughout the mission (i.e., during winter darkness and spring sunshine), and is not stationary in System III as proposed by previous studies \citep{08fletcher_poles, 09baines_pole}.  Indeed, we observe a westward rotation of the hexagon by $\approx30^\circ$ between March 2007 and April 2013, which is qualitatively consistent with the drift rates estimated from the recent reflectivity analysis of \citet{14sanchez}.

\item \textbf{Evolving stratospheric vortices: } Counter to the predictions of radiative climate models, which expect the rate of change of temperatures ($\pderiv{T}{t}$) to vary relatively smoothly with latitude, we find that the most intense stratospheric northern warming and southern cooling is localised within the polar regions (i.e., within 15$^\circ$ of both poles), with changes of approximately $\pm5$ K/year over the mission duration.  The peak warming in the north is in the 0.5-1.0 mbar region, whereas the peak cooling in the south is in the 1-3 mbar region.  In the south, this coincides with the `summer stratospheric hood' of bright emission \citep{05orton, 08fletcher_poles}, which has cooled significantly over time (by approximately 35 K) so that the sharp meridional temperature gradient near 75$^\circ$S has disappeared.  The strong southern stratospheric methane emission of 2004 is not apparent in 2014, suggesting that the summer stratospheric vortex is dissipating.  As of 2014, there is no northern counterpart to the bright southern summer emission, although stratospheric temperatures do rise smoothly from 75$^\circ$N to 90$^\circ$N at 1 mbar, and are continuing to warm during northern spring.  Cassini is still awaiting the onset of a sharp thermal gradient near 75$^\circ$N which would signify the presence of a northern summer stratospheric vortex.  Such a region of enhanced emission should be present before northern summer solstice in May 2017 ($L_s=90^\circ$) if Saturn's temperatures follow the same pattern as the last Saturnian year \citep[e.g.,][]{89gezari}.

\item \textbf{Hydrocarbons within the polar vortices: }  Acetylene and ethane abundances are at their minimum near the 75$^\circ$S boundary in the stratosphere, and increase towards the south pole.  The hydrocarbon latitudinal gradient is most dramatic in acetylene (1-mbar abundance varying from 0.2 ppm at 75$^\circ$S to 0.6 ppm at 90$^\circ$S at the start of the mission) and more gradual for ethane (6 ppm to 11 ppm from 75$^\circ$ to 90$^\circ$S).  Acetylene, with the shorter photochemical lifetime and steeper vertical gradient (as a fraction of the abundance), appears to have a greater response to the stratospheric circulation than ethane. These peak abundances may have initially been generated by subsidence from the source regions at microbar pressures.  At the north pole, hydrocarbon emissions were barely discernible at the start of the mission but now exhibit rising abundances from 75-90$^\circ$N, consistent with the warming trend observed.  However, the summertime spike in hydrocarbon abundances within 5$^\circ$ of the south pole is not yet mirrored in the springtime north.

\item \textbf{Atmospheric response lags: } The measured atmospheric response lags behind the changing insolation (i.e., temperature minima lag behind winter solstice in the north; maxima lag behind summer solstice in the south), and this lag generally increases with pressure due to the longer radiative timescales in the deeper atmosphere.  CIRS clearly detected the north polar minima between 2008-2010 (6-8 years after winter solstice, approximately one season), but the south polar maxima occurred close to or before the start of the CIRS dataset in 2004 (2 years after summer solstice).  These lags are entirely consistent with the radiative climate predictions of \citet{08greathouse_agu}, \citet{12friedson} and \citet{14guerlet}.  Intriguingly, the 5-mbar lag in the southern stratosphere appears to be longer than at 100 mbar, suggesting that the atmosphere continues warming for longer in the mid-stratosphere than the upper troposphere (i.e., a longer radiative time constant at this altitude than expected).  The cause of this is unknown, but may be indicative of an additional component to the radiative budget (e.g., the presence of stratospheric aerosols).  Such an effect is not reproduced in the radiative models to date.

\item \textbf{Vertical circulation from temperatures: }  Radiative climate models do not reproduce the increase in stratospheric temperatures poleward of 75-80$^\circ$N, nor the gradients in stratospheric temperature near 75$^\circ$S and 87$^\circ$S.  To explain these significant departures, one must invoke either additional opacity contributions to the radiative budget (e.g., UV-absorbent stratospheric aerosols, entrained within the polar vortices and locally warming the atmosphere in daylight), or explain the temperature contrasts in terms of stratospheric circulation causing adiabatic heating and cooling.  We use the retrieved $\pderiv{T}{t}$ and an estimate of the deviation from radiatively-forced temperatures to estimate the vertical velocities, and require subsidence at $w\approx-0.15$ mm/s to warm the northern stratosphere, and upwelling at $w\approx+0.11$ mm/s to cool the southern stratosphere.  As expected, we find these regions of upwelling and subsidence to be localised within the stratospheric polar regions poleward of $\pm75^\circ$.  Velocities tend towards zero in the deeper atmosphere where temperature variability is negligible, except within the polar cyclonic vortices themselves: at 100 mbar we predict velocities of $w\approx-0.02$ mm/s in the NPC and $w\approx+0.02$ mm/s in the SPC.  

\item \textbf{Vertical circulation from hydrocarbons: }  Both 1-mbar acetylene and ethane abundances have been increasing in the north polar region and decreasing in the south polar region over the duration of the Cassini observations.  North polar ethane enhancements seem largest within the polar vortex ($0.45\pm0.1$ ppm/year poleward of 75$^\circ$N, $0.10\pm0.03$ ppm/yr equatorward), with a sharp boundary in between.   This boundary is not visible in the north polar acetylene changes, which increase from $0.005\pm0.005$ ppm/yr at 75$^\circ$N to $0.030\pm0.003$ ppm/yr at 85$^\circ$N.  South polar compositional changes appear localised within 5$^\circ$ of the pole, where acetylene depletion rises from 0.01 ppm/yr to $0.030\pm0.005$ ppm/yr, and ethane depletion rises from 0.1 ppm/yr to $0.35\pm0.1$ ppm/yr.  Diffusive photochemistry models \citep{05moses_sat,07moses} predict negligible variability at 1 mbar, and this behaviour is instead closely tracking the expected variations in hydrocarbon production at much lower microbar pressures.  Furthermore, chemical models do not predict the sharp boundaries in $\pderiv{q}{t}$ evident in our results.  This suggests that atmospheric circulation is dominating the rate of change of composition with time.  The continuity equation is used to reveal the vertical velocities (or vertical gradient changes) required to explain the compositional results, estimating north/south polar subsidence/upwelling at 0.15 mm/s from acetylene (consistent with the derivations from the temperature field).  However, the ethane $\pderiv{q}{t}$ requires larger velocities ($\pm0.6$ mm/s) in the polar regions, suggesting that modifications to the vertical hydrocarbon gradients are also occurring.


\end{enumerate}

Saturn's polar atmosphere exhibits the most extreme seasonal variability found on the giant planet, and this study has quantified the magnitudes and directions of those changes over a third of a Saturnian year.  We eagerly await the results from the final phases of the Cassini mission approaching the northern summer solstice, particularly the development of strong gradients of stratospheric emission signifying the onset of the warm northern stratospheric hood.  Only through the longevity and unique vantage point offered by a robust orbiting spacecraft have we been able to study the polar environment during every season, from winter to spring, and summer to autumn, providing our most comprehensive understanding of the seasonal evolution of a giant planet to date.


\section*{Acknowledgments}
The analysis presented in this paper would not have been possible without the tireless efforts of the CIRS operations and calibration team, who were responsible for the design of the imaging sequences, instrument commands and other vital operational tasks.  Fletcher was supported by a Royal Society Research Fellowship at the University of Oxford.  We are indebted to S. Guerlet and T. Greathouse for their willingness to share the outputs of their radiative climate models for comparison with our thermal retrievals; J. Moses for sharing her 2D photochemical model output for comparison with the hydrocarbon distributions; and A. Antu\~{n}ano for sharing cloud-tracked wind velocities at high polar latitudes.  The UK authors acknowledge the support of the Science and Technology Facilities Council (STFC).   Orton was supported by grants from NASA to the Jet Propulsion Laboratory, California Institute of Technology.

\bibliographystyle{elsarticle-harv}
\bibliography{references_master}







\end{document}